\documentclass[pdftex,twocolumn,epjc3]{svjour3}          

\RequirePackage[T1]{fontenc}

\smartqed  
\usepackage{amsmath}
\usepackage{hhline}

\RequirePackage{graphicx}
\RequirePackage{mathptmx}      
\RequirePackage{flushend}
\RequirePackage[numbers,sort&compress]{natbib}
\RequirePackage[colorlinks,citecolor=blue,urlcolor=blue,linkcolor=blue]{hyperref}
\RequirePackage{lineno,hyperref}
\journalname{Eur. Phys. J. C}

\begin{document}

\title{The BDX-MINI detector for Light Dark Matter search at JLab} 

\author{M. Battaglieri\thanksref{GEaddress,JLABaddress}
        \and P. Bisio\thanksref{GEaddress,UNIGEaddress}
        \and M. Bond\'i\thanksref{e0,e1,GEaddress}
        \and A. Celentano\thanksref{GEaddress}
        \and P.L. Cole \thanksref{LAMARaddress}
        \and M. De Napoli\thanksref{CTaddress}
        \and R. De Vita\thanksref{GEaddress}
        \and L. Marsicano\thanksref{GEaddress} 
        \and G. Ottonello\thanksref{GEaddress}
        \and F. Parodi\thanksref{GEaddress}
        \and N. Randazzo\thanksref{CTaddress}
        \and E.S. Smith\thanksref{JLABaddress}
        \and D. Snowden-Ifft\thanksref{OCCaddress}
        \and M. Spreafico\thanksref{GEaddress,UNIGEaddress}
        \and T. Whitlatch\thanksref{JLABaddress}
        \and M.H. Wood\thanksref{CANISIUSaddress}
}

\thankstext{e0}{Corresponding author}
\thankstext{e1}{Mariangela.Bondi@ge.infn.it}

\institute{INFN - Sezione di Genova, Via Dodecaneso 33, I-16146 Genova, Italy\label{GEaddress}
          \and
          Jefferson Lab, Newport News, VA 23606, USA\label{JLABaddress}
          \and
          Universit\`a degli Studi di Genova, 16146 Genova, Italy\label{UNIGEaddress}
          \and
          Lamar University, 4400 MLK Boulevard, P.O. Box 10046, Beaumont, Texas 77710, USA\label{LAMARaddress}
          \and
          INFN - Sezione di Catania, Via S. Sofia 64, I-95125 Catania, Italy\label{CTaddress}
          \and
          Canisius College, Buffalo, NY, 14208, USA\label{CANISIUSaddress}
          \and
          Occidental College, Los Angeles, CA 90041, USA \label{OCCaddress}
}

\date{Received: date / Accepted: date}

\maketitle

\begin{abstract}

This paper describes the design and performance of a compact detector, BDX-MINI, that incorporates all features of a concept that optimized the detection of light dark matter produced by electrons in a beam dump. It represents a reduced version of the future BDX experiment expected to run at JLAB.  
BDX-MINI was exposed to penetrating particles produced by a 2.176\,GeV electron beam incident on the beam dump of Hall A at Jefferson Lab. The detector consists of 30.5\,kg of PbWO$_4$ crystals with sufficient material following the beam dump to eliminate all known particles except neutrinos. The crystals are read out using silicon photomultipliers. Completely surrounding the detector are a passive layer of tungsten and two active scintillator veto systems, which are also read out using silicon photomultipliers. The design was validated and the performance of the robust detector was shown to be stable during a six month period during which the detector was operated with minimal access. 

\end{abstract}


\section{Introduction}
BDX is a Beam Dump eXperiment searching for Light Dark Matter particles in the MeV-GeV mass range produced by the interaction of the Jefferson Lab (JLab) multi-GeV, high-intensity electron-beam with the experimental Hall-A beam dump~\cite{Bondi17}. BDX will have the requisite sensitivity to explore a entirely new physics regime (“Dark Sector”), where light dark matter (LDM) is the lightest stable state. LDM particles $\chi$ are charged under a new U(1)$_D$ broken symmetry, whose mediator is a massive vector boson called  $A^\prime$ or ``dark photon''. The dark photon can be kinetically mixed with the Standard Model (SM) hypercharge field, resulting in SM-DM interaction. We refer the reader to the recent review works~\cite{Battaglieri:2017aum,Filippi:2020kii,Fabbrichesi:2020wbt} for a comprehensive discussion of the LDM theory and the corresponding ongoing experimental effort.

In the BDX experiment, a high-current (up to 150 $\mu$A) $e^-$ beam accelerated to energy up to 11 Gev by CEBAF,  hits the experimental Hall-A Al/water beam dump~\cite{753410}, producing, together with a large number of SM particles,  a ``dark beam''  of $\chi$ particles. LDM is produced in the beam dump via two main mechanisms:  $A^\prime$-strahlung~\cite{Liu:2017htz,Gninenko:2017yus}, which is conceptually akin to bremsstrahlung for SM,  and $e^+e^-$ resonant annihilation~\cite{Marsicano:2018glj}. 
The beam dump is heavily shielded with 20 m of concrete, iron and dirt, acting as a filter for almost all SM particles. Only weakly interacting particles (SM $\nu$ and LDM $\chi$) propagate through the shielding to the BDX detector that is designed to identify rare interactions. 
The BDX detector is  a homogeneous electromagnetic calorimeter, surrounded by a dual-layer veto system to reject the cosmogenic background~\cite{Marsicano:2020ztz}. The $\chi$ interaction with the atomic electrons in the calorimeter results in a high-energy scattered $e^-$ that can be easily detected in the BDX detector.
The scattering is identified by detecting a high-energy electromagnetic shower with no associated activity in the surrounding veto systems. Considering the low probability of LDM production and detection, beam-dump experiments require intense beams and a large accumulated charge.   
With  more than $10^{22}$ electrons-on-target (EOT) expected in one year of running, BDX will cover altogether unexplored regions of the parameter space of LDM coupling versus mass, thereby exceeding, by up to two orders of magnitude, the discovery potential of existing and planned experiments.
The BDX experiment was approved with the highest scientific rating by the JLab Program Advisory Committee in 2018~\cite{Batta}. While preparing for the design and construction of a new subterranean hall downstream of the existing Hall-A beam dump, the BDX collaboration fabricated and deployed a reduced version of the full detector, BDX-MINI, performing a first, preliminary search for LDM at lower beam energy.
We have fabricated and deployed a reduced version of the BDX detector, BDX-MINI, performing a first, and as such, a preliminary search for LDM at low beam energy. 
BDX-MINI ran in spring/summer 2020 collecting a charge of $\simeq 4\times10^{21}$ EOT.
This paper describes the BDX-MINI detector design and construction, and its performance compared to Monte Carlo simulations. The document is organized as follows. In Section~\ref{sec:setup}, we present the experimental set-up  describing in detail the BDX-MINI detector components. In Section~\ref{sec:MC}, we describe the Monte Carlo simulations used to study the detector response to the LDM signal and background. Finally, in Section~\ref{sec:results} we report on the performance of the BDX-MINI detector and its individual components during production and dedicated calibration runs taken in 2020 and 2019, respectively.

\section{The BDX-MINI experiment at JLab: experimental setup   \label{sec:setup}  }

\subsection {Detector location}
\begin{figure}[t]
\centering
\includegraphics[width=2.9in]{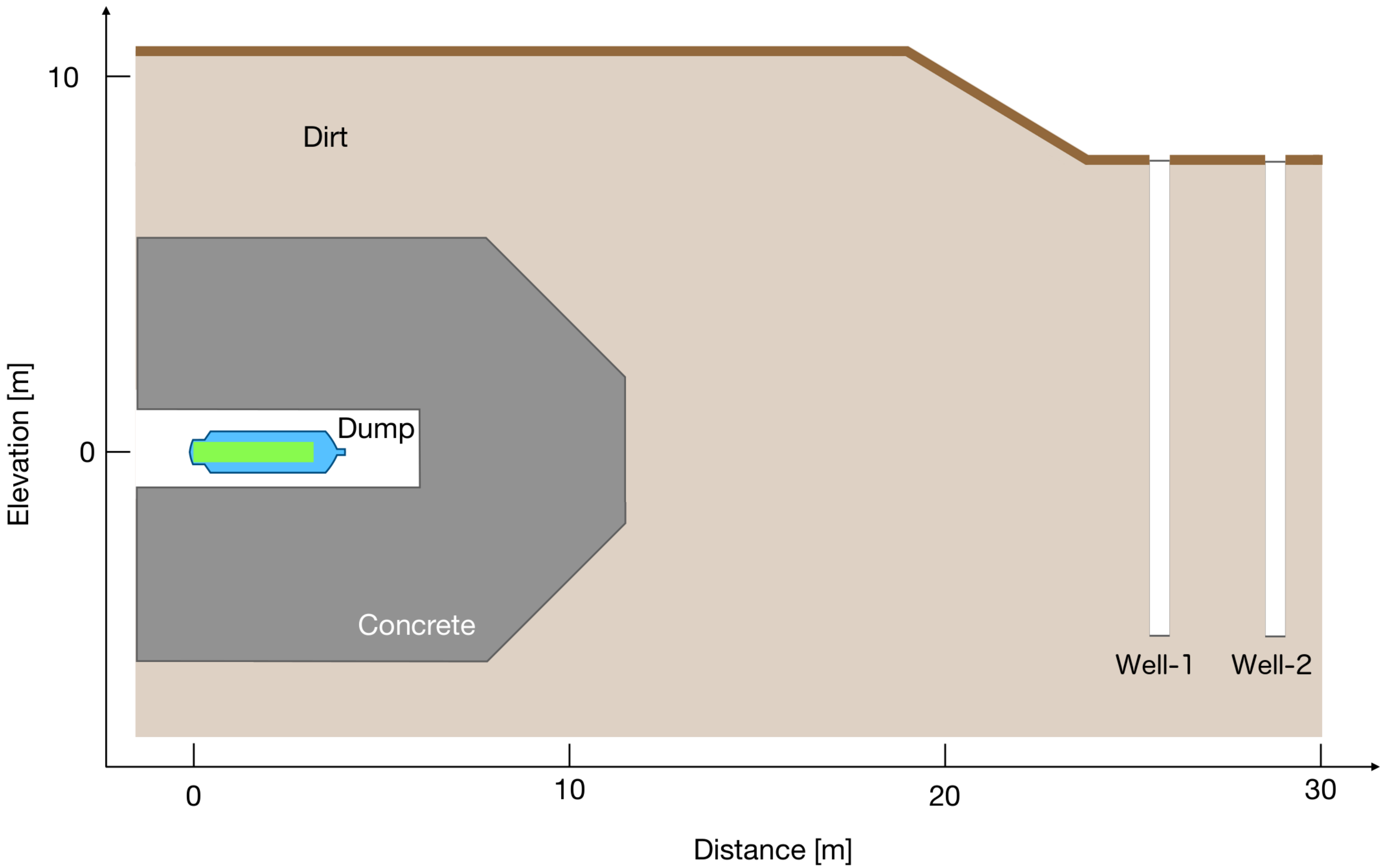}
\caption{Schematic representation of the location of the wells relative to the Hall-A beamdump. From left to right, the Hall-A aluminum-water beamdump (blue-green), the concrete beam-vault walls (gray), the dirt (brown), and the two vertical pipes. The detector was located in the well closest to the accelerator, Well-1. (Color online)}
 \label{fig:pipes_location}
\end{figure}

The detector was installed  26 m downstream of the Hall-A beam dump at JLab, inside a well (Well-1) at beamline height.
The detector was shielded from the  background produced  in the beam dump by 5.4 m of concrete and 14.2 m of dirt, shown schematically in Fig.\,\ref{fig:pipes_location}. BDX-MINI accumulated data during a period of six months between 2019 and 2020. During most of the time, Hall A received one-pass beam from the accelerator (2.176~GeV) and all muons generated in the beam dump ranged out before reaching the detector. In a previous paper, we described the physical layout and measurements of the muon flux produced by 10.6~GeV electrons incident on the beam dump \cite{Battaglieri:2019ciw}. The present measurements use a new detector, which is deployed in the well closest to the accelerator. 

\begin{figure}[t]
\centering
\includegraphics[width=.45\textwidth]{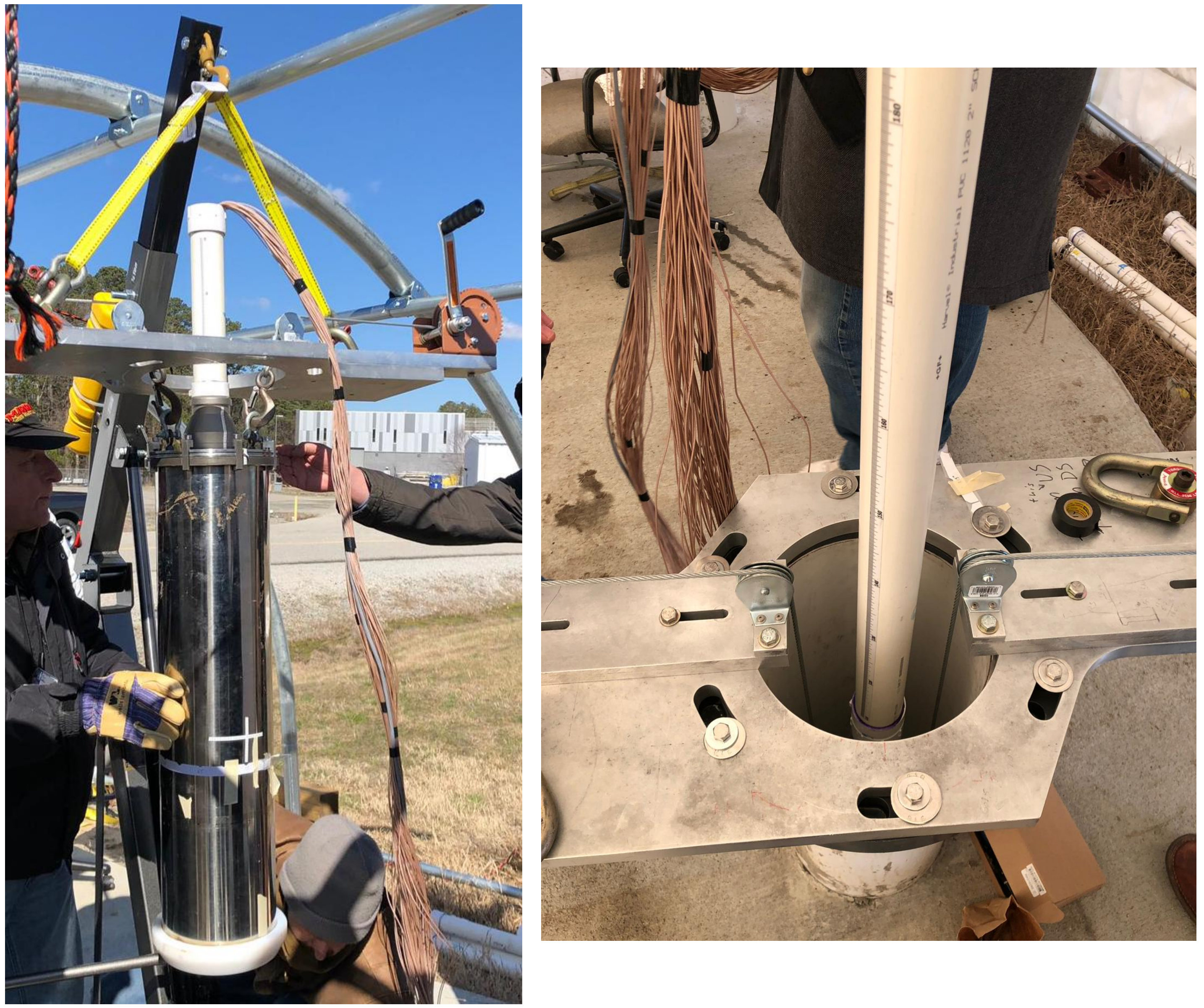}
\caption{Some pictures of BDX-Mini detector positioning phases. Left Panel shows the BDX-mini detector inside the stainless-steel watertight cylindrical vessel. Right Panel shows the detector lowering phase in the pipe: a PVC segment with the attached tape is cleary visible.}
 \label{fig:vesselBDXMINI}
\end{figure}

The detector was situated inside a 20-cm diameter stainless-steel watertight cylindrical vessel, shown in Fig.~\ref{fig:vesselBDXMINI}, covered by steel lids on top and bottom, supported by steel cables that were used to move it up and down using a hand winch. The electronic cable connections to the detector were routed through a 5\,cm diameter  PVC pipe that was firmly attached to the top of the vessel. The PVC pipe came in six 122\,cm-sections that could be screwed together (or unscrewed) as the detector was hoisted up and down inside the well. The height of the detector was determined using a tape measure attached to the individual PVC segments with registration marks to ensure reproducibility as they were connected and disconnected. The location of the detector at beam height was checked by measuring a non-stretch fishing line lowered down with a plumb bob to the top of the vessel.
The systematic error associated to the procedure was estimated to be $\Delta Y_{Pos}$\,=\,$\pm$\,5\,cm. 

The readout electronics and the DAQ system were mounted on a dedicated electronic rack at ground level near the entrance to the well. To protect the BDX-MINI from the weather, the entire experimental setup was situated within a sturdy field tent that covered both wells, the electronic equipment, and power breakers. The environment was further conditioned using a portable air conditioning unit to maintain a suitable temperature and humidity.

\subsection{BDX-MINI detector}

The BDX-MINI is a complete detector package that incorporates, on a reduced scale, all the detector elements that are envisioned for the BDX experiment. The compact electromagnetic calorimeter (ECal) composed of PbWO$_{4}$ crystals ($\sim$\,4\,x\,10$^{-3}$\,m$^{3}$ total volume), is surrounded by layers of hermetic passive and active vetoes. The innermost layer consists of passive tungsten shielding (WS), shaped as a prism with an octagonal base, with tungsten lids on top and bottom. The middle (Inner Veto, IV) and the outer (Outer Veto, OV) layers are active detectors, made of plastic scintillators. The IV matches the WS octagonal shape, while the OV is a cylinder. Each of them are sealed, on top and on the bottom, by plastic scintillator lids. The whole detector is inserted in a stainless steel vessel for protection and handling.
\begin{figure*}[htbp!] 
\centering
\includegraphics[width=\textwidth]{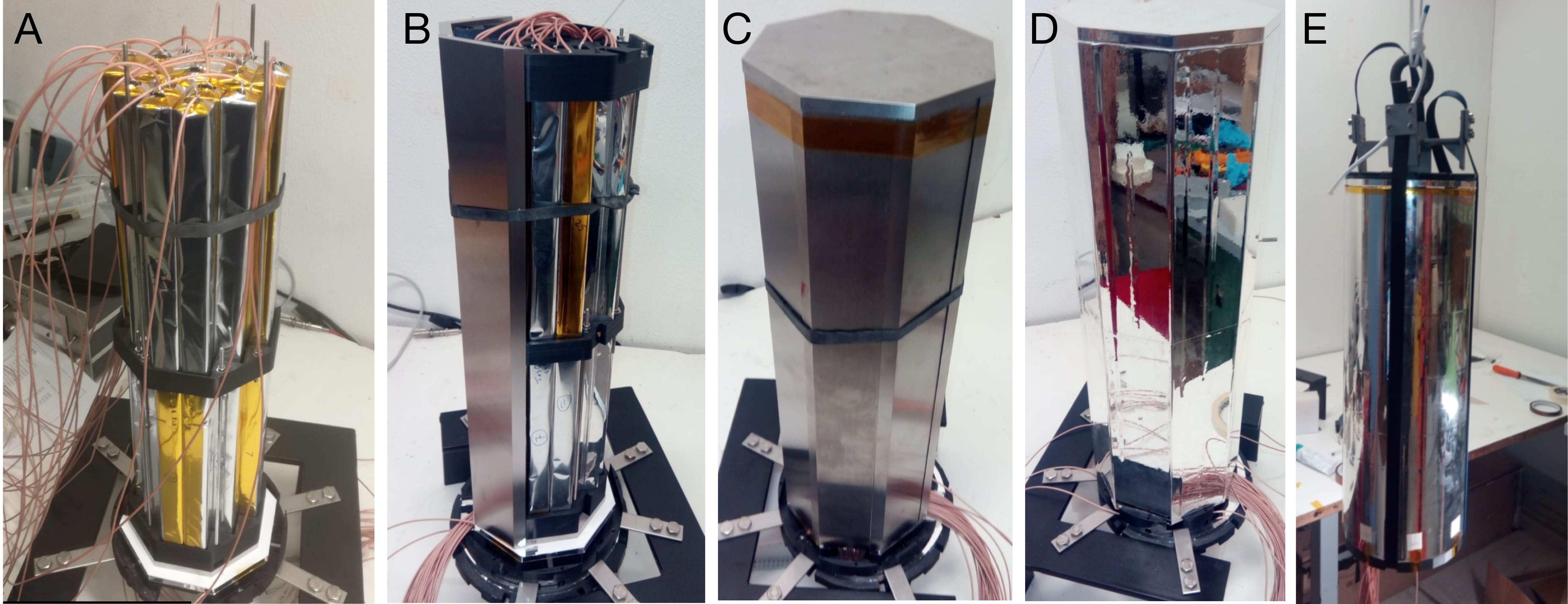}
 \caption{The different components of BDX-MINI are showed while they were assembled. Panel A shows full assembled ECal, Panel B and C the tungsten shielding, Panel D the inner veto, and Panel E the outer veto.}
 \label{fig:BDXMINI_detector}
\end{figure*}

Figure~\ref{fig:BDXMINI_detector} shows the ECal, the W-shielding, and the two active vetoes during the detector construction. The BDX-MINI components are described in detail in the following subsections. 

\subsubsection{Electromagnetic Calorimeter}
The BDX-MINI ECal is composed of  60 parallelepiped PbWO$_{4}$ crystals. Two types of lead-tungstate crystals are employed: 32 15\,x\,15\,x\,200\,mm$^{3}$ crystals, produced by the SICCAS\footnote{Shanghai Institute of Ceramics, Chinese Academy of Science} company and  formerly used as spares in CLAS12 - Forward Tagger detector (FT)~\cite{Acker:2020brf}, and 28 20\,x\,20\,x\,200\,mm$^{3}$ crystals, produced by the BTCP\footnote{Bogoroditsk Technical Chemical Plant} company,  used as spares in the PANDA-ECal~\cite{Erni:2008uqa}. To reduce the number of readout channels, the FT crystals were glued in pairs along the 15\,x\,200\,mm$^{2}$ face to obtain 16 30\,x\,15\,x\,200\,mm$^{3}$ compound crystals. Each of the resulting 44 crystals were then wrapped in a reflecting aluminized Mylar foil. 

The BDX-MINI ECal is assembled from two modules (as shown in Fig.~\ref{fig:BDXMINI_detector}, Panel\,A), each composed of 14 PANDA and 8 compound FT crystals. 
\begin{figure}[b] 
\centering
\includegraphics[width=2.5in]{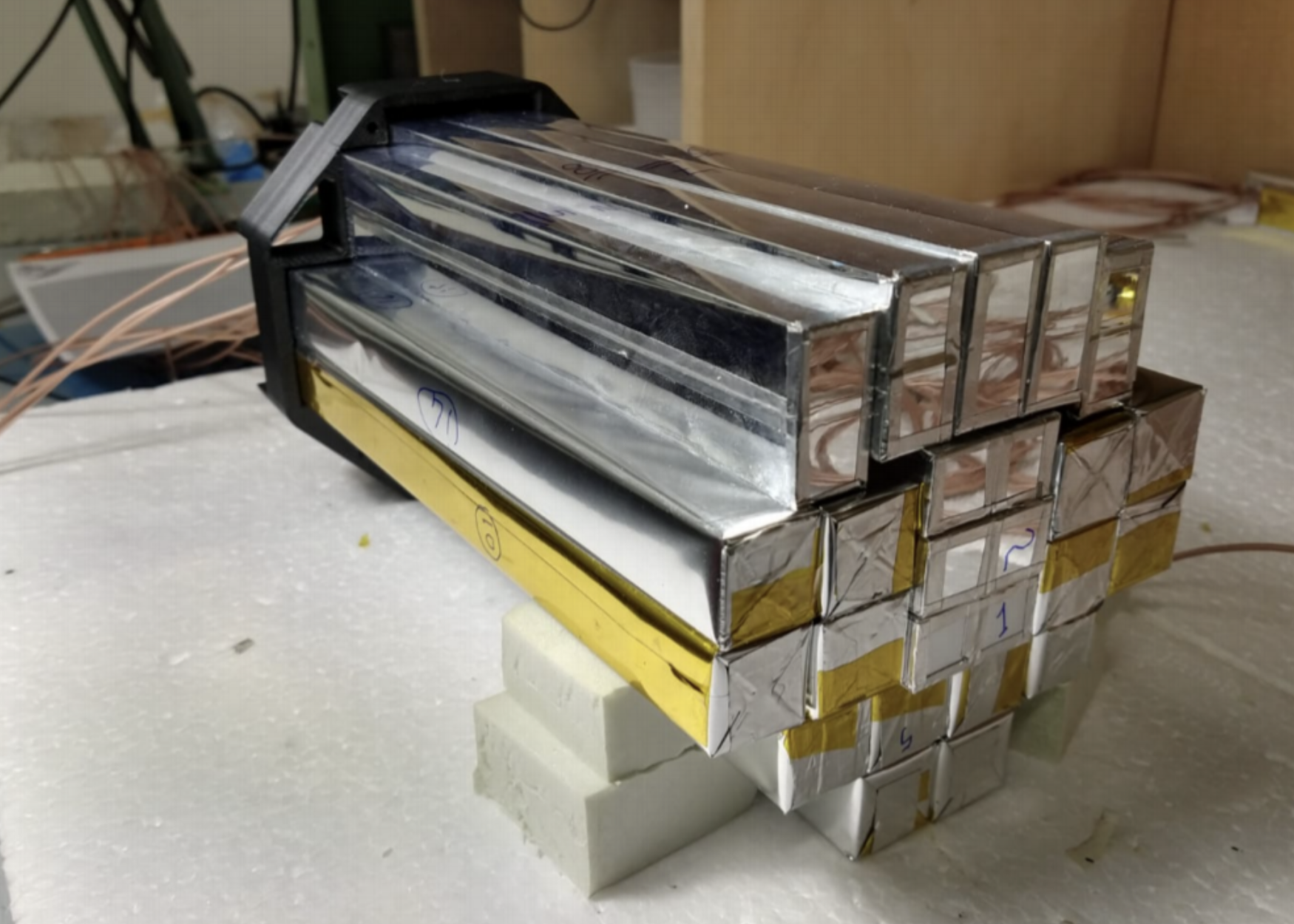}
 \caption{Lower module of the BDX-MINI ECal.}
 \label{fig:ECAL_module_BDXMINI}
\end{figure}
A 3D-printed plastic mechanical support holds the crystals in place. The two modules are mounted vertically, resulting in an approximately cylindrical shape 40\,cm long and 11.5~cm in diameter (see Fig.~\ref{fig:ECAL_module_BDXMINI}).

The BDX-MINI ECal uses 6\,x\,6\,mm$^{2}$ Hamamatsu MPPCs (S13360-6025PE) to read out the PbWO$_{4}$ scintillation light.
The sensor is placed in the middle of the 20\,x\,20\,mm$^{2}$ face of each PANDA crystal and optically coupled with optical grease. Likewise, the sensor is optically coupled on the junction between the two glued crystals of the 30\,x\,15\,mm$^{2}$ face of the FT compound crystal. 

\subsubsection{Tungsten Shielding}
The ECal is enclosed in a 0.8\,cm-thick tungsten shield (Fig.~\ref{fig:BDXMINI_detector}\,C), shaped as an octagonal prism, with length $l$ = 45\,cm and base side 5\,cm. It is composed of 10 separate parts: two octagonal plates form the prism upper and lower bases, while the lateral surface is made of 8 plates. In the lower octagonal plate a small open slot allows the passage of the ECal signal cables. The purpose of the passive tungsten layer between the vetoes and the ECal is to protect the active vetoes from electromagnetic showers produced by $\chi - e^-$ interactions that may accidentally self-veto the interaction.

\subsubsection{Inner Veto}
The IV is composed of EJ200 plastic scintillators, 0.8\,cm thick, forming an hermetic octagonal prism with a 6.2\,cm base side  and 49.4\,cm in length (Fig.~\ref{fig:BDXMINI_detector}\,D). Two octagonal scintillators are used for the upper and lower prism bases. Machined into each of them is a spiral groove, which contains a Wavelength Shifter (WLS) fiber used to collect and transfer the light to a SiPM. The lower base has a small open slot to allow the cables to come out. The lateral surface of the prism is composed of 8 bars coupled to each other with optical glue in order to obtain an optically continuous piece (see Fig.~\ref{fig:veto_BDXMINI}\,B). In each bar, four WLS fibers are inserted into a linear groove running down the middle and parallel to the long side of the plastic, transporting the scintillation light to a SiPM light sensor. Both the IV and the OV (described in the next section) use 3x3\,mm$^{2}$ Hamamatsu S13360-3075CS SiPMs. The bias voltage and amplifiers boards are the same ones used in the ECal, but with a lower gain factor.

\subsubsection{Outer Veto}

\begin{figure}[t] 
\centering
\includegraphics[width=.45\textwidth]{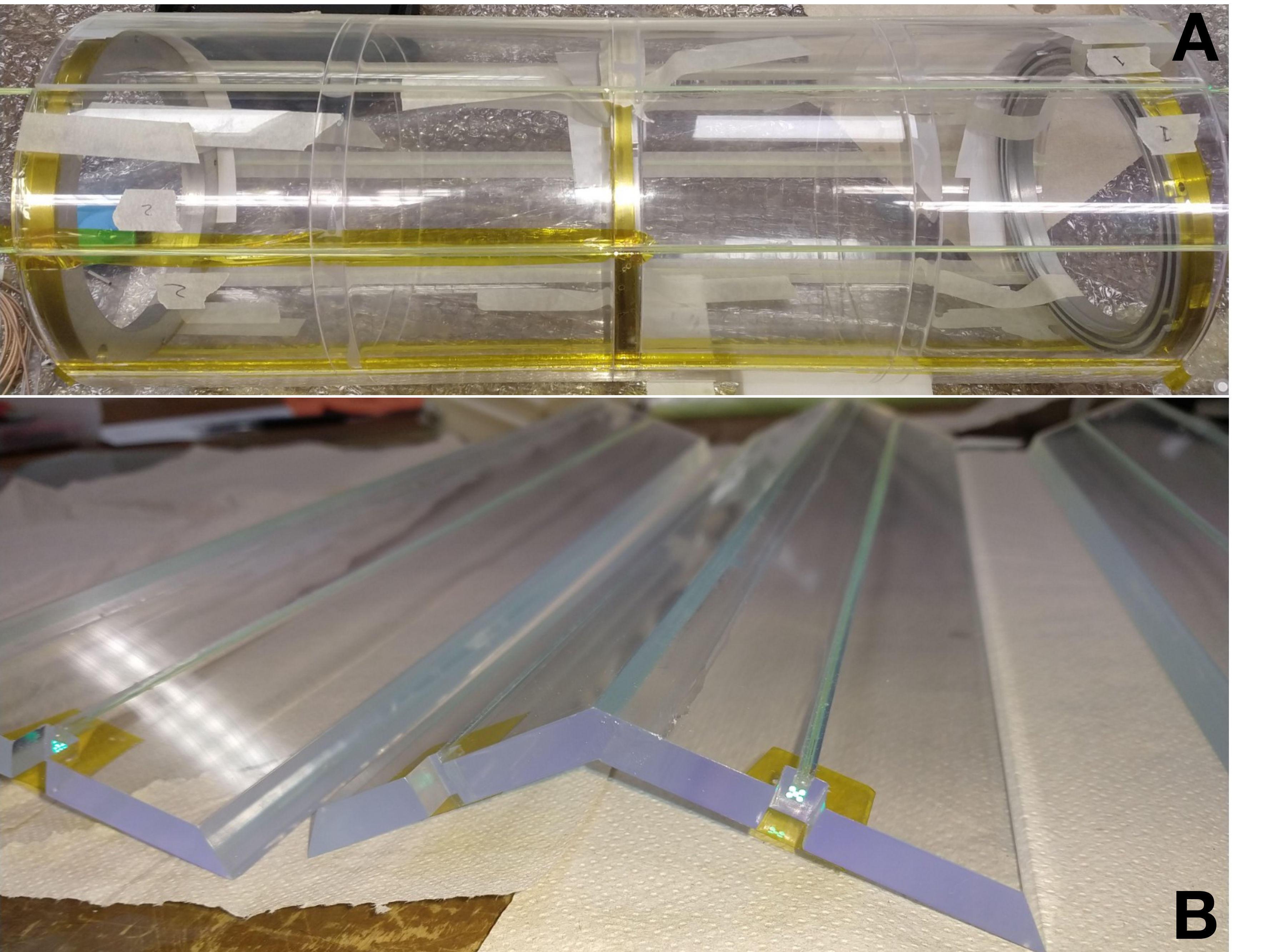}
 \caption{Panel A: OV cylinder picture where the grooves for the WLS fibers are clearly visible. Panel B: Scintillator bars forming the IV lateral surface.}
 \label{fig:veto_BDXMINI}
\end{figure}

The OV, made up of 0.8\,cm thick EJ200 plastic scintillators, is cylindrical in shape with radius of 9.7\,cm and 53.0\,cm in height.
It is formed by two round caps and a single cylindrical tube (Fig.~\ref{fig:BDXMINI_detector}\,E). Light produced in each cap is extracted by a wavelength-shifting fiber inserted into a spiral groove on the paddle's surface and coupled to a SiPM (Fig.~\ref{fig:veto_BDXMINI}\,A). As for both the IV and the WS, the lower OV cap also has an open slot to extract the cables. 
The cylinder features eight equally spaced grooves, running along the length of the scintillator and  each containing four WLS fibers. In this configuration, both the IV
octagonal prism and the OV cylinder have a redundant readout of eight SiPMs each.

\subsection{Front-end electronics}

Each SiPM is connected via an 8-m long coaxial cable to a custom front-end circuit that provides the bias voltage to the photosensor and amplifies the corresponding signal. The SiPM cathode is connected to the circuit input via the coaxial cable inner conductor, while the anode is connected to ground via the cable outer shield. We found that the noise associated with this connection scheme was sufficiently low to distinguish the individual photo-electrons signals for each and every one of the SiPMs. Signal amplification is obtained by means of a custom transimpedance amplifier, with a gain factor of 100 for ECal, 80 for IV octagon and OV cylinder, and 44 for IV and OV caps. It provides two equal output signals through a built-in 1:1 passive resistor divider following the last amplification stage. The bias voltage is provided by a custom-designed board, which uses a 5\,V input tunable DC-DC converter~\cite{Cele_17}. 
All the front-end circuits are mounted in a rack 
outside the well, which also contained the data acquisition electronics and computer. 

\subsection{Data acquisition and trigger system}

As mentioned before, each detector signal is amplified by a custom circuit, which provides two equal outputs of the signal. The first copy of the signal is sent to a leading-edge discriminator (CAEN v895), whose digital output is sent to a programmable logic board (CAEN FPGA v1495) implementing a custom trigger logic. A threshold of $\sim10$ MeV is implemented for the crystals, while for veto signals a threshold of few photo-electrons is used.
The second copy is fed to a Flash Amplitude-to-Digital converter. Signals from the inner and outer veto photodetectors were processed with a 2 V, 14 bit, 500 MHz module (CAEN FADC v1730), while those from the crystals were processed with a board featuring a lower 250 MHz sampling rate (CAEN FADC v1725). To guarantee synchronization, the clock was generated from the first board and distributed to the others through a daisy-chain setup. 

The trigger conditions implemented for the experiment are reported in Tab.~\ref{tab:trigger}. For each bit, the table gives the average rate corresponding to the nominal detector operations (2.176 GeV beam or no beam). The first two triggers, i.e. the ``OR'' of all top/bottom crystal signals, were used for the LDM measurement. In parallel to these, other trigger equations were implemented for monitoring, calibration and debugging. Each trigger bit could be prescaled by a user-programmable factor, from 1 to 16384. The global trigger condition is the union of all individual trigger bits after pre\-scale. For each trigger, all the FADC raw waveforms were written to the disk without further processing. In order to monitor the rates in the detector, as well as the trigger rate and the livetime, individual scalers were implemented in the FPGA firmware and regularly read through the slow controls system, as discussed in the next section.

We used the standard JLab ``CEBAF Online Data Acquisition'' (CODA) software to handle the readout system~\cite{CODA}, and software from a recent measurement, with minor modifications, to allow for an increased number of boards~\cite{Battaglieri:2019ciw}. 

\begin{table}[tpb]
\centering
\begin{tabular}{rll}
\hline
\textbf{Trg \#} & \textbf{Definition} & \textbf{Typ. rate}\\
\hhline{===}
0 & Top Crystals OR & 1.6 Hz \\ 
1 & Bottom Crystals OR& 1.6 Hz\\
2 &IV/OV Top Caps coincidence & 1.5 Hz\\
3 &IV/OV Bottom Caps coincidence & 0.9 Hz\\
4 &IV Top/Bottom Caps coincidence & 0.1 Hz\\
\hline
\end{tabular}
\caption{\label{tab:trigger} Summary of trigger bits implemented for the BDX-MINI detector.}
\end{table}

\subsection{Slow controls}

A custom, EPICS-based slow-controls system was developed for the BDX-MINI setup~\cite{Dalesio:1994qp}. Several software Input Output Controllers (IOCs) are implemented, each associated to a different set of Process Variables (PVs) in the experimental setup. The system is integrated with the main JLab slow-controls system, in order to access accelerator-related PVs, for example reporting the beam current and energy. A custom graphical interface was developed to interact with the PVs. During data-taking runs, the values of all PVs are periodically recorded to the data stream.

The temperature and the humidity at the detector location was monitored by two probes (model Adafruit 3251, embedding a Si7021 sensor) installed inside the watertight cylindrical vessel. Both probes are read by an Arduino Uno board, connected to the main slow-controls system and interfacing with a specific IOC.
Likewise, at the experimental area inside the tent, where the readout electronics was housed, the temperature and humidity were monitored with an AKCP sensorProbe device, interfaced to another IOC.
Finally, a third IOC has been developed to read the FPGA scalers and configure the trigger bit prescale values.

\subsection{Online monitoring system\label{par:online_monitoring}}

To monitor the detector performance during data taking and to rapidly identify and thereby correct any problems as quickly as possible, various online monitoring tools were developed.
The count rate for each detector channel, for each trigger bit, and for the total trigger rate were displayed on a custom, EPICS-based, graphical user interface (GUI). Trigger bits rates were shown both before and after prescaling. This GUI also showed the temperature and humidity at the detector location and in the experimental area. The time evolution of these quantities was also monitored using the EPICS StripTool program.

During data taking, an online reconstruction system monitored both single-event FADC waveforms and accumulated observables, including the spectrum of the energy deposited in each crystal, and the total energy deposited in the two BDX-MINI modules. The online system interfaced with the DAQ through the JLab ``Event Transfer System (ET)''~\cite{ET}, that provides a very fast and robust method of transferring events between different processes. The online reconstruction was based on the same JANA-based libraries that are used in the offline reconstruction. Single-event FADC waveforms and histograms were displayed via the RootSpy system~\cite{RootSpy}.

\section{Monte Carlo simulations  \label{sec:MC} }

We used Monte Carlo simulations in order to study the interaction of cosmic rays, beam-induced muons produced in the beam dump, and beam-induced neutrinos with the BDX-MINI detector. We performed different simulations, using both the FLUKA~\cite{Bohlen:2014buj,Ferrari:2005zk} and the Geant4~\cite{Agostinelli:2002hh} codes. The interaction of neutrinos with the detector was simulated using the GENIE program~\cite{Andreopoulos:2009rq,Andreopoulos:2015wxa}. 

\subsection{Detector geometry and response}

\begin{figure}[t]
\centering
\includegraphics[width=.35\textwidth,height=8cm]{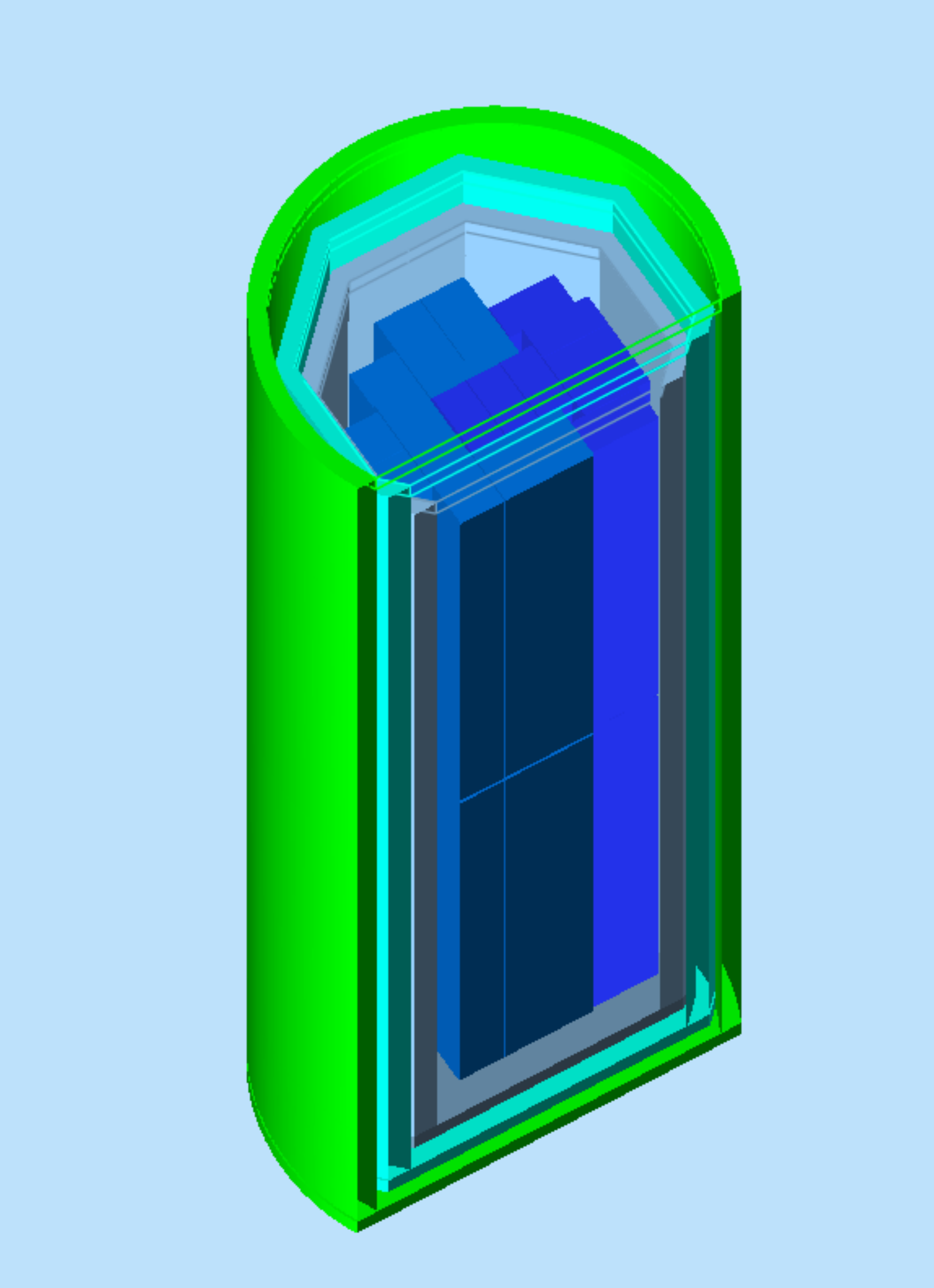}
\caption{\label{fig:gemcGeometry} Geant4 implementation of the BDX-MINI detector - for better visibility, the inner/outer veto top caps and the tungsten top shielding are shown in wireframe style. See text for details. (Color online).}
\end{figure}

The geometry, the materials and the response of the BDX-MINI detector were implemented in Geant4 via the GEMC interface \cite{Ung16}. Figure~\ref{fig:gemcGeometry} shows the detector geometry as implemented in the Geant4 geometry. The PbWO$_4$ crystals are shown in light blue (PANDA crystals) and dark blue (FT crystals), surrounded by the tungsten shielding (gray). Outside the shielding, the inner veto (cyan) and the outer veto (green) are shown. The inner veto and outer veto top caps are show in yellow and orange, respectively, in wireframe style.
The simulation does not include the photosensors and the inner/outer veto grooves hosting the WLS fibers -- the effect of these on the veto response was effectively included in the corresponding response parameterization, as discussed in the following.   

\begin{figure}[t]
\centering
\includegraphics[width=.45\textwidth]{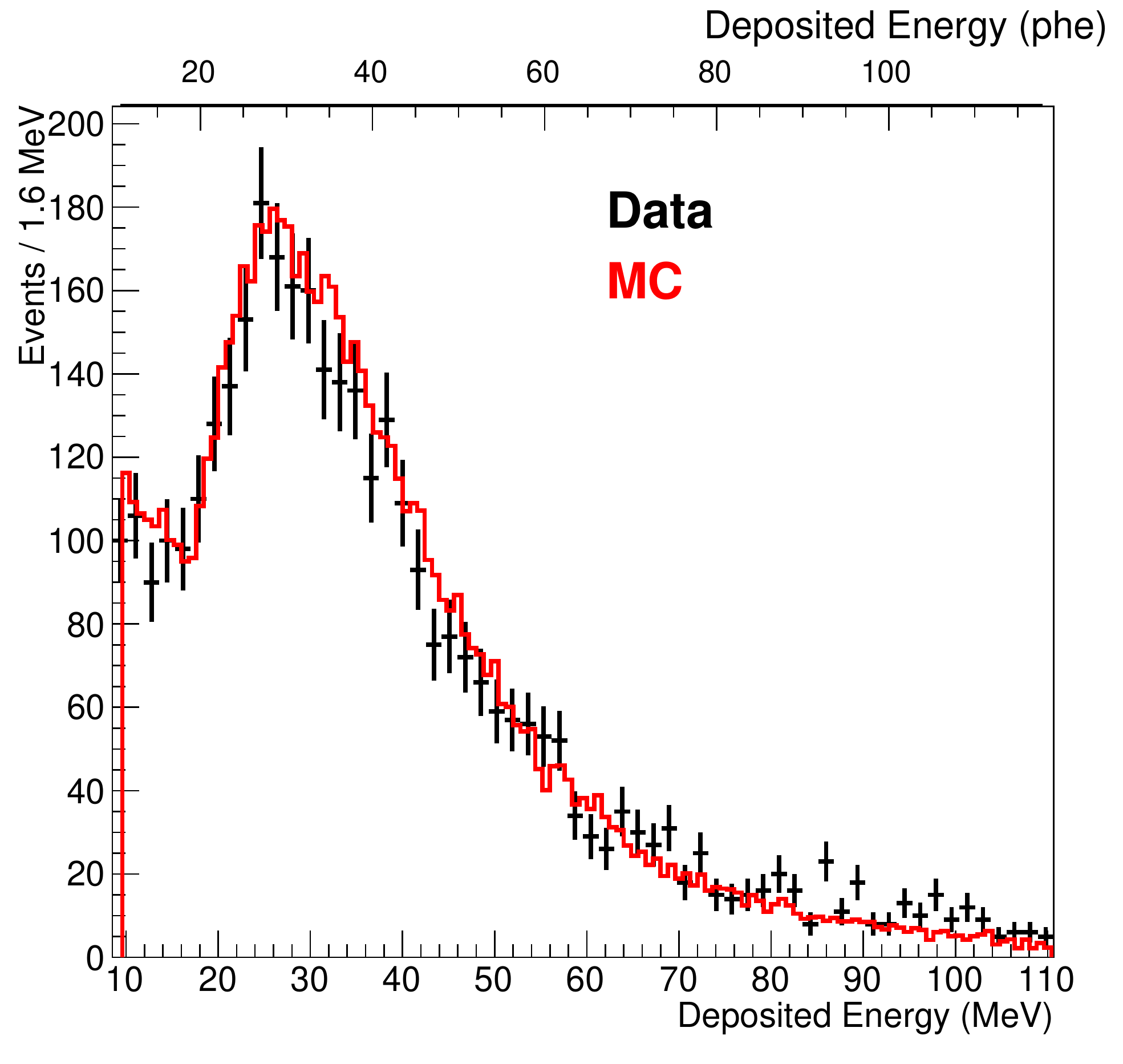}
\caption{PbWO$_4$ crystal cosmic-rays energy spectrum: data are shown in black, while simulations results are reported in red. (Color online).}
\label{fig:PbWO4spectrum} 
\end{figure}

The description of the PbWO$_4$ crystal response includes: a light yield (LY) of 310 photons per MeV deposited, a decay time driven by two time constants (64$\%$ for the 6.8~ns fast component, 36$\%$ of the 33.4~ns slow component) and the SiPM nominal photon detection efficiency (PDE) of 22$\%$. Due to the limited size of the crystals, no light attenuation length has been included in the code. The same parameters were used for both types of crystals. Systematic differences among FT and PANDA crystals, as well as differences among the two batches, were absorbed in individual calibration constants in data analysis.
The comparison between the simulated cosmic muon energy deposition spectrum and the measured one is shown in Fig.~\ref{fig:PbWO4spectrum}. 
The absolute normalization of the simulated distribution was scaled to the measured one. The two distributions are in very good agreement, confirming the validity of our parameterization in Monte Carlo of the crystal response to muons.
Furthermore, we note that this marks the first use of SiPMs in high-energy electromagnetic calorimetry with PbWO$_4$ crystals. The result here reported demonstrates that the use of a 6x6~mm$^2$ SiPM with 25~$\mu$m pixel size coupled to a PbWO$_4$ crystal results to a light yield of ${\sim}$1~phe/MeV. The use of four of these SiPMs per crystal could achieve a statistical resolution term of ${\sim}{2\%}/\sqrt{E(\mathrm{GeV})}$, avoiding saturation effects up to 50-100~GeV.

A detailed description of the veto geometry and photoelectron response was implemented in  the BDX-MINI simulation framework. Both the IV and OV caps and lateral surfaces were described as separate scintillator pieces. For each scintillator, the deposited energy of a hit was converted to photoelectrons. 
As described above, the lateral surfaces of the IV and OV consist of optically continuous large scintillator pieces, with multiple SiPM readout. When a charged particle hits them, scintillation light travels to the WLS fibers, where it is collected and transported to SiPMs. As a result, all photo-detectors produce a signal with an
amplitude depending on the position of the WLS fibers relative to the hit point on the scintillator.
In particular, there are two main effects in the  light transmission to each WLS fiber. First, light transmitted in the plastic will be attenuated in accordance with the distance traveled.
Given the dimension of the two components, this effect cannot be neglected. Secondly, a large fraction of light is absorbed whenever it crosses one of the grooves containing the WLS fibers. This translates to a discretized light reduction depending on the number of grooves that the scintillation light crosses while propagating from the hit point to a particular sensor. To account for this behavior, an effective formula (for both the IV and OV) was implemented in the MC, describing the light attenuation factor for the $i$-th SiPM depending on the energy deposition position:

\begin{equation}
    a_i(\phi,z) = F_G^{N_G} \, (1-Z_L) \, e^{-\frac{\phi}{\phi_{att}}}\, +\, \\ F_G^{7 -N_G} \, (1-Z_R) \, e^{-\frac{2 \pi - \phi}{\phi_{att}}}.
\label{eq:att}
\end{equation}
\begin{figure}[t]
\centering
\includegraphics[width=.52\textwidth]{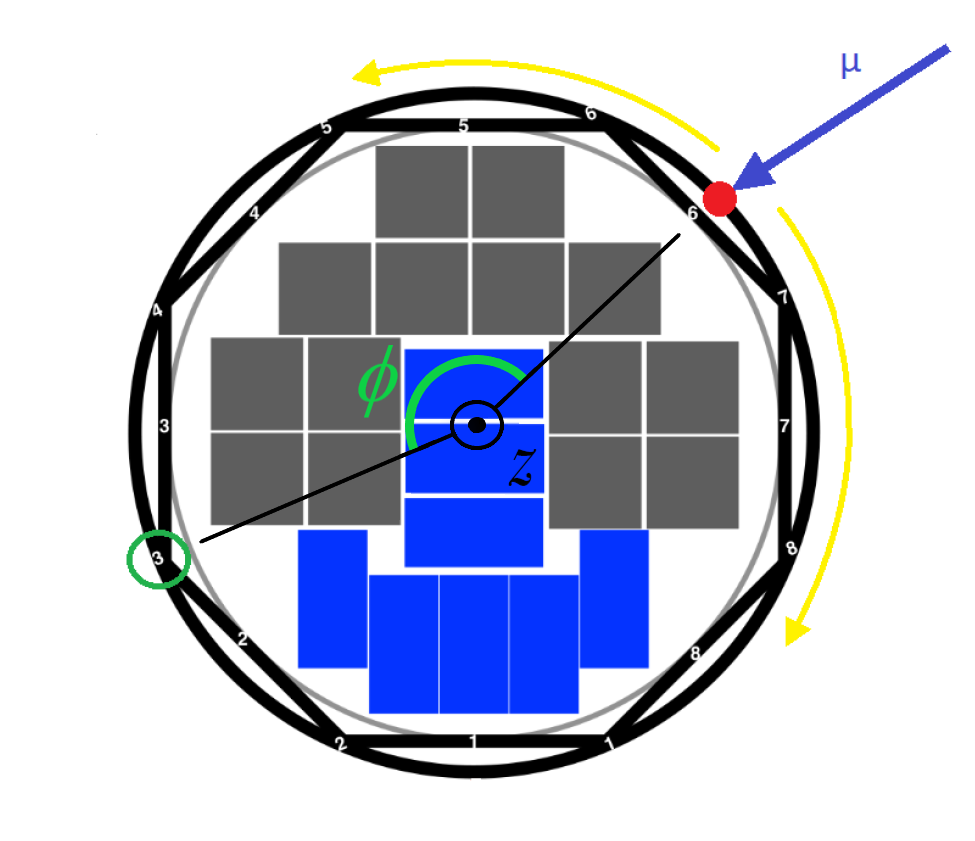}
\caption{ Scheme showing the light transmission in the OV lateral surface. When a charged particle hits the veto, scintillation light is emitted from the impact point (red dot in the picture). The generated light then travels through the scintillator both clockwise and counterclockwise. As an example, to reach the WLS fiber coupled to SiPM 3 (in the green circle) light has to cross the grooves containing the other fibers (photons traveling clockwise cross 4 grooves, while light moving counterclockwise crosses 3 grooves). (Color online).}
\label{fig:scheme}
\end{figure}
Here  $\phi$  is the azimuthal angle between the  charged particle hit position and the axis defined by the WLS fiber of the $i$-th SiPM and $z$ is the hit vertical position, being $z=0$ defined as the lower edge of the IV and OV (see Fig.~\ref{fig:scheme}). $F_G$ is the parameter describing the fraction of scintillation photons absorbed when a single groove is crossed, $N_G$ is the number of grooves light crosses to reach the $i$-th fiber traveling counterclockwise looking the scintillator from the top and $\phi_{att}$ is  the parameter defining the angular attenuation length. Finally,  $Z_L$ and $Z_R$ are the factors describing the attenuation light undergoes as it travels vertically along the scintillator:

\begin{align}
Z_L =& \left( 1-e^{-\frac{z}{z_{att}}} \right) \cos({\phi}/2)\mathcal{H}(\pi-\phi)\\
Z_R =& \left( 1-e^{-\frac{z}{z_{att}}} \right) \cos((2\pi-\phi)/2)\mathcal{H}(\phi-\pi),
\end{align}
where $z_{att}$ is the light attenuation length along $z$, and $\mathcal{H}$ is the Heaviside step function.
The two terms in Eq.~\ref{eq:att} represent the two different paths light can follow. Since the IV and OV lateral surfaces are optically continuous, the light generated by a particle reaches the WLS fiber of a given SiPM traveling both clockwise and counterclockwise.

All the parameters in Eq.~\ref{eq:att}  were tuned by comparing cosmic-ray data to MC. 
We selected several muon trajectories that hit the lateral surfaces of the vetoes using the coincidences between the crystals and veto caps. 
The different parameters of Eq.~\ref{eq:att} were then adjusted by comparing the peak position of the charge distribution of the SiPMs from the data to MC.  The values of $F_G$ and $\phi_{att}$ where obtained by selecting muons trajectories hitting the vetoes with fixed $z$, while trajectories crossing the vetoes at different heights were used to evaluate the value of  $z_{att}$.
This procedure was employed separately for the IV and the OV in order to obtain two different set of parameters. The values $F_G = 0.6$ and $z_{att} = 100$ cm where found to describe the light attenuation in both the IV and OV. However, two different values of the attenuation length along $\phi$ were obtained: $\phi_{att\,OV} = \frac{7}{9} \pi$,  $\phi_{att\,IV} = \frac{2}{3}\pi$.

\subsection{Beam-induced muons}
Muons produced in the beam dump and reaching the detector are a very useful calibration source. There was a short period in 2019 when the accelerator delivered 10.38\,GeV e$^{-}$-beam to Hall~A and the corresponding high flux of muons that traversed the detector were used for calibration. 
We note that during the LDM production period all muons produced in the beam dump ranged out before reaching the BDX-MINI detector.

In order to simulate the production, propagation, and detector interaction of muons produced by the interaction of the primary $e^-$ with the Hall-A beam dump, we adopted the same procedure detailed in Ref.~\cite{Battaglieri:2019ciw}. We used the existing configuration of the Hall-A beam dump geometry and materials implemented in FLUKA-2011.2c.5 by the Jefferson Lab Radiation Control Department \cite{Kharas}. The input card used to run FLUKA includes all physics processes and a tuned set of biasing weights to speed up the running time while preserving accuracy. FLUKA was used to generate muons in the beam dump and propagate them to a plane near the detector. From there, Geant4 was used to track particles all the way up to the detector, and to simulate its response.

Using this procedure, we verified that particles produced by the 2.176\,GeV electron beam, except neutrinos, are fully absorbed by the concrete vault and the dirt between the  dump and the detector, thus not contributing at all to the background rate in the detector. At the same time, Monte Carlo simulations predicted a sizable interaction rate due to beam-related muons when the primary beam energy was set to 10.38\,GeV. The signal corresponding to the propagation of these particles through the BDX-MINI detector was used for calibration, alignment, and efficiency studies, as discussed in Section~\ref{sec:results}.

\subsection{Cosmic rays \label{sec:MC_cosmic_rays}}
Our characterization of the cosmogenic backgrounds in the BDX-MINI was determined empirically from data taken during beam-off periods. Cosmic rays, however, are exploited also during beam-on periods to monitor the detector stability and to cross-check the energy calibrations obtained with beam-induced muons. We did so by identifying specific muon trajectories, selected using a combination of energy thresholds in the crystals and topological selections in the inner and outer veto systems, to make data to MC comparisons. Therefore, we required a realistic generator of cosmic muons.

The cosmic muon energy spectrum and angular distribution reported in Ref.~\cite{Appleton:1971kz} were implemented directly in Geant4, using the algorithm originally developed for the study of cosmogenic backgrounds in the Modane underground laboratory~\cite{Kluck:2013fum}. Cosmic particles were generated in a fiducial volume large enough to contain the detector and a careful normalization has been performed to correctly take into account the crossing on the lateral sides. Particles found crossing the fiducial volume where then projected out to production vertices outside the dirt. In this way, the simulation correctly accounted for the effects of energy loss, multiple scattering, and secondary particles production in the overburden above the detector.

\subsection{Neutrinos}

Although the beam-related neutrino background corresponds to a much lower interaction rate in the detector compared to the dominant cosmogenic one, it represents an \textit{irreducible} contribution to the LDM measurement. Therefore, we performed a detailed Monte Carlo calculation of this background.

\begin{figure}[t]
    \centering
    \includegraphics[width=.48\textwidth]{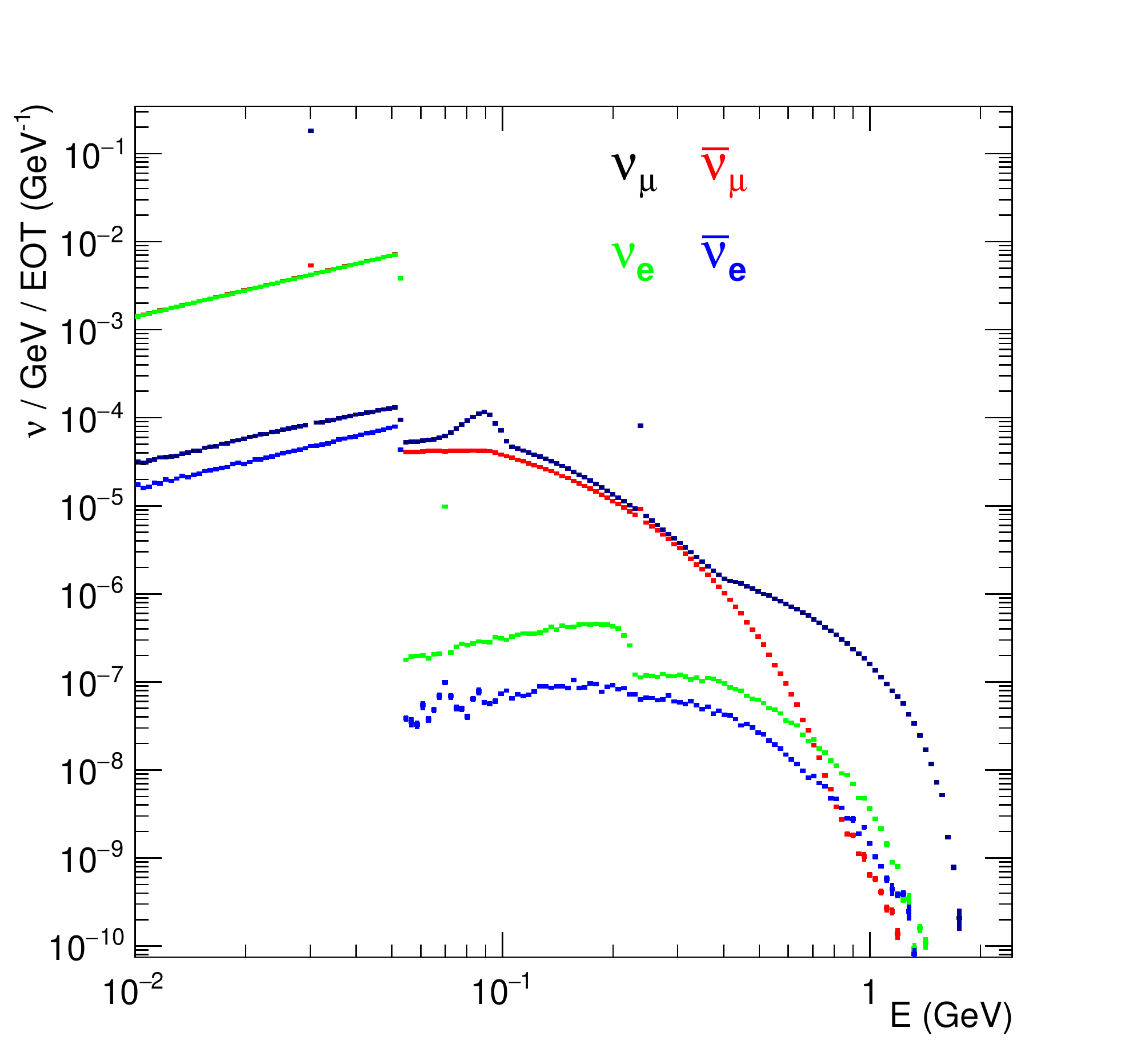}
    \caption{Neutrino differential flux, per EOT, of neutrinos produced by the interaction of the CEBAF 2.176 GeV $e^-$ beam with the Hall-A beam dump. The flux is scored on a plane located 0.5 m upstream the BDX-MINI detector.
    Each color correspond to a different neutrino topology, as reported in the Figure. (Color online).}
    \label{fig:nuFlux}
\end{figure}

The FLUKA calculation sampled the neutrino flux using the same setup adopted for the simulation of beam-induced muons. We computed the flux on a scoring plane located 0.5~m upstream the BDX-MINI detector, perpendicular to the primary $e^-$ beam direction.
In particular, to preserve correlations among kinematic variables, we sampled the 3D distribution $\Phi_\nu(E_\nu,\cos(\theta_\nu),R^2_\nu)$, where $E_\nu$, $\cos(\theta_\nu)$, $R^2_\nu$ are, respectively, the neutrino energy, the cosine of the neutrino momentum angle with respect to the primary $e^-$ beam direction, and the neutrino radial impact point at the scoring plane distance from the primary $e^-$ axis. The $\Phi_\nu$ distribution was sampled independently for each neutrino topology ($\nu_e$, $\overline{\nu}_e$, $\nu_{\mu}$, $\overline{\nu}_{\mu}$). The energy distribution of neutrinos produced by the 2.176 GeV $e^-$ beam, integrated over the other variables, is shown in Fig.~\ref{fig:nuFlux}.

Neutrino interactions with the detector elements and with the surrounding materials were simulated using the GENIE generator, v. 3.0.6~\cite{Andreopoulos:2009rq,Andreopoulos:2015wxa}. A custom application was developed using the GENIE libraries, in order to use the input neutrino flux $\Phi_\nu$ discussed previously and the same BDX-MINI detector geometry implemented in Geant4 -- this was imported in GENIE using the GDML format. \\The \texttt{G18\_02a\_00\_000} tune was used for neutrino cross-section data. 

\begin{figure}[t]
    \centering
    \includegraphics[width=.47\textwidth]{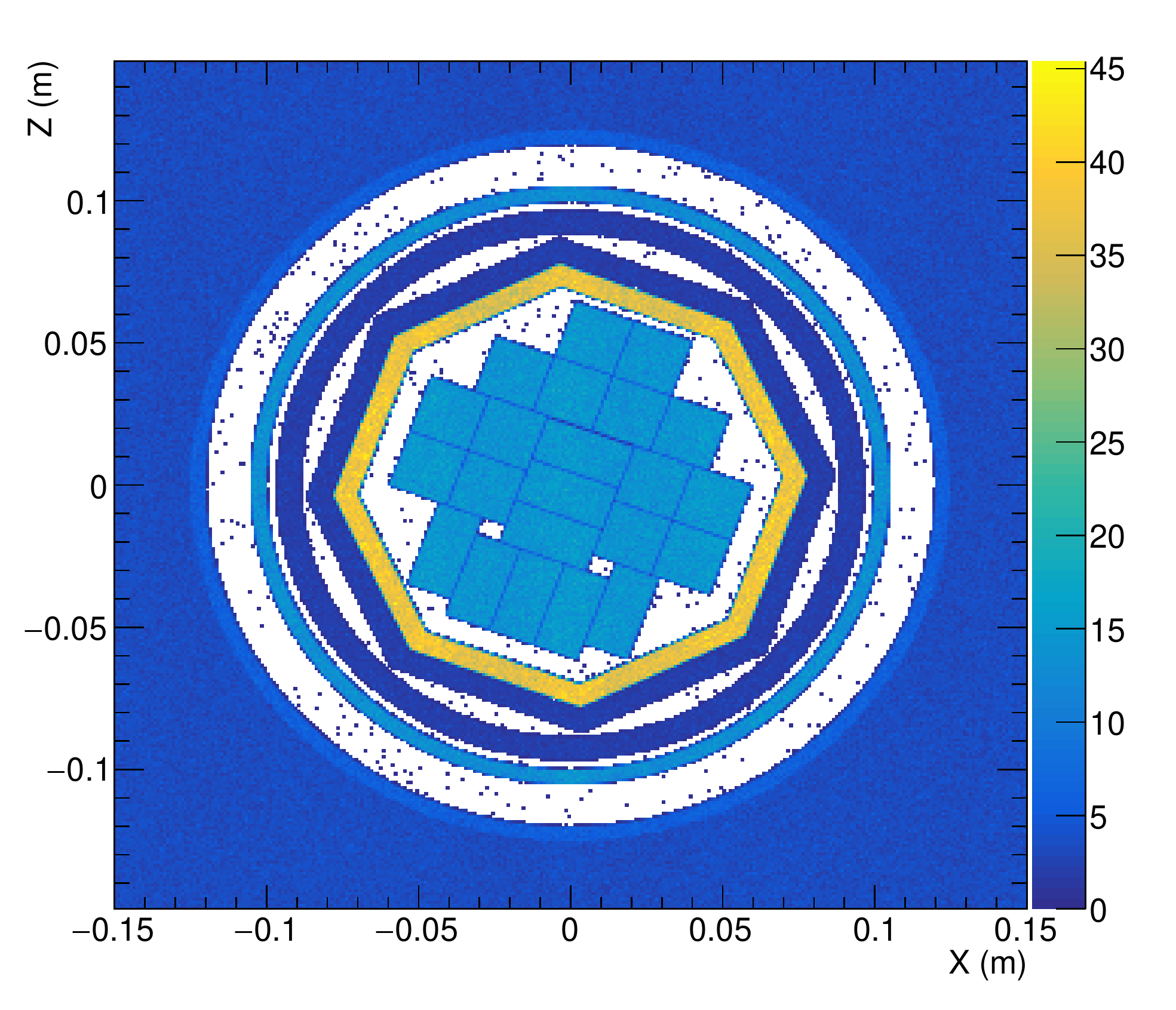}
    \caption{Distribution of neutrino interaction vertices (in arbitrary units) in the BDX-MINI detector as obtained from the GENIE simulation. The color scale reflects the interaction probability and therefore the density of detector materials. Cyan: PbWO$_4$ crystals, yellow: tungsten shielding, dark blue: IV and OV: OV, light blue: stainless steel vessel, white: air between vessel and PCV pipe, blue: PCV pipe and surrounding dirt. (Color online).
    }
    \label{fig:nuVertex}
\end{figure}

The GENIE simulation output is a list of unweighted neutrino events, together with the corresponding normalization (i.e. the number of EOT that would result in the same number of neutrino interactions). For each event, GENIE provides the interaction vertex position and the list of secondary particles. To speed-up the simulation, only events within a 40-cm radius fiducial sphere centered at the detector are considered. As an example, the obtained vertex distribution of $\nu$ events is shown in Fig.~\ref{fig:nuVertex}. The shape of the detector is clearly visible, with neutrinos vertices distributed according to the different materials densities.

\begin{figure}[t]
    \centering
    \includegraphics[width=.45\textwidth]{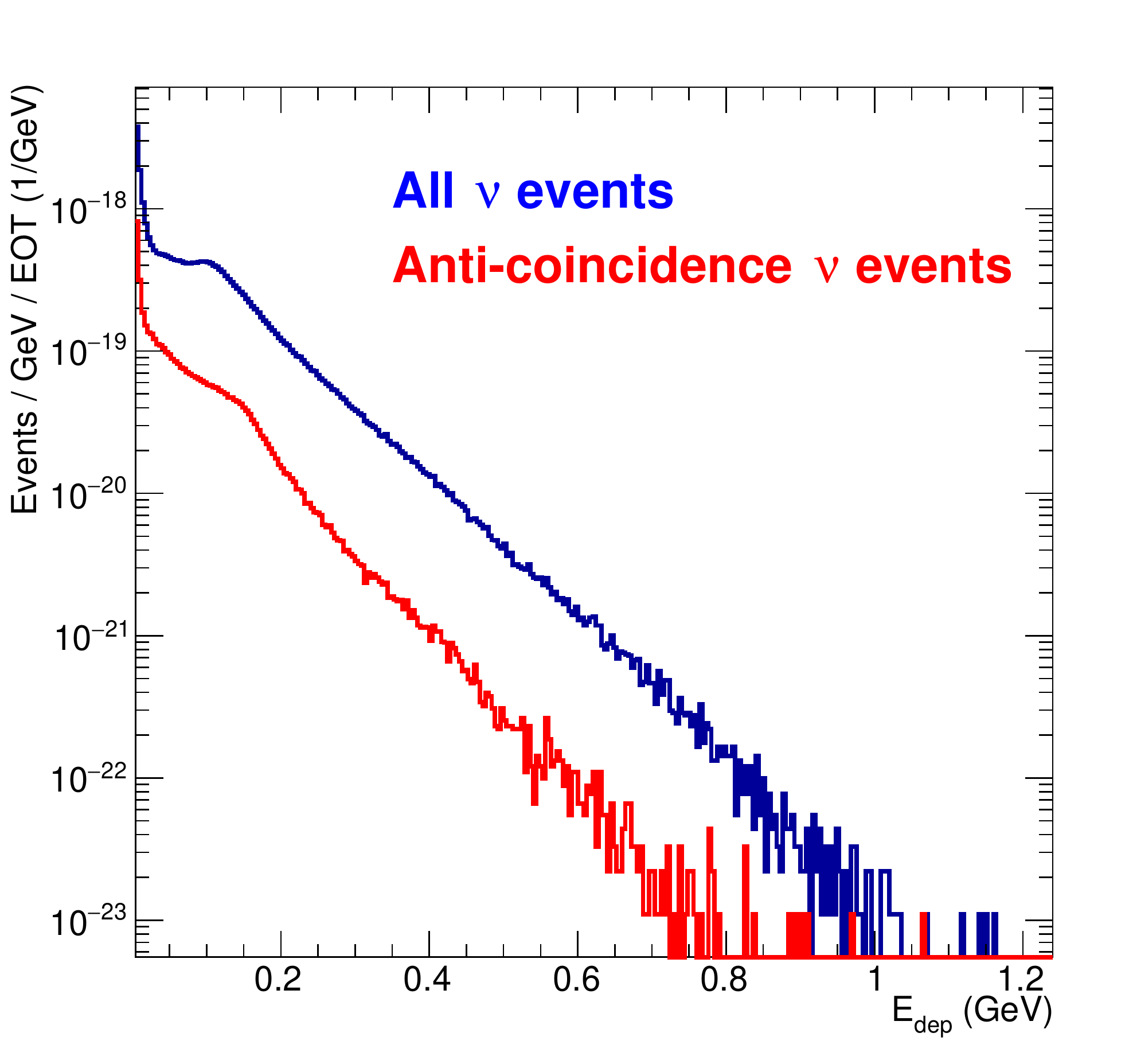}
    \caption{Neutrino energy deposition spectrum in the BDX-MINI crystals, for all events (blue) and for events with no signal in the surrounding veto (red). (Color online).}
    \label{fig:nuEdep}
\end{figure}

The GENIE events were then used as input to the Geant4-based simulation of the BDX-MINI detector in order to study the corresponding response. Figure~\ref{fig:nuEdep} shows the neutrino energy deposition spectrum, per EOT, in the BDX-MINI crystals, comparing all events (black) with events where there is no activity in the surrounding veto systems (red). The latter anti-coincidence condition is defined as the absence of any veto SiPM signal with charge greater than 5 phe\footnote{This value is similar to the final threshold envisioned for the inner/outer veto systems in the LDM analysis.}. The total number of events for an energy threshold of 0.2 GeV is $1.1 \cdot 10^{-20}/\mathrm{EOT}$ and the number of anti-coincidence events is $1.1 \cdot 10^{-21}/\mathrm{EOT}$.\footnote{The relatively high number of self-vetoed events is largely due to $\nu_\mu$ that produce penetrating muons during charged-current interactions.} This corresponds to a total number of anti-coincidence background neutrino events of $\sim 4.4$ for the accumulated charge of $4\cdot10^{21}$ EOT that was accumulated during the BDX-MINI measurement.

\section{Results}\label{sec:results}
BDX-MINI accumulated data for approximately six months in 2020. During most of this time, Hall A received 2.176\,GeV e$^{-}$ beam with currents ranging from few up to 150\,$\mu$A. At 2.176\,GeV, no beam-related SM particles, except neutrinos, are expected to reach the BDX-MINI detector. A few special runs were taken in late 2019 at 10.381\,GeV with currents from 5 to 35\,$\mu$A. At this energy we observe a large flux of muons produced in the dump \cite{Battaglieri:2019ciw}. Muons traversing the detector were used for energy calibration and detector characterization, as discussed in Section\,\ref{EnergyCalib}.  

\begin{figure}[htpb!]
    \centering
   \includegraphics[width=.45\textwidth]{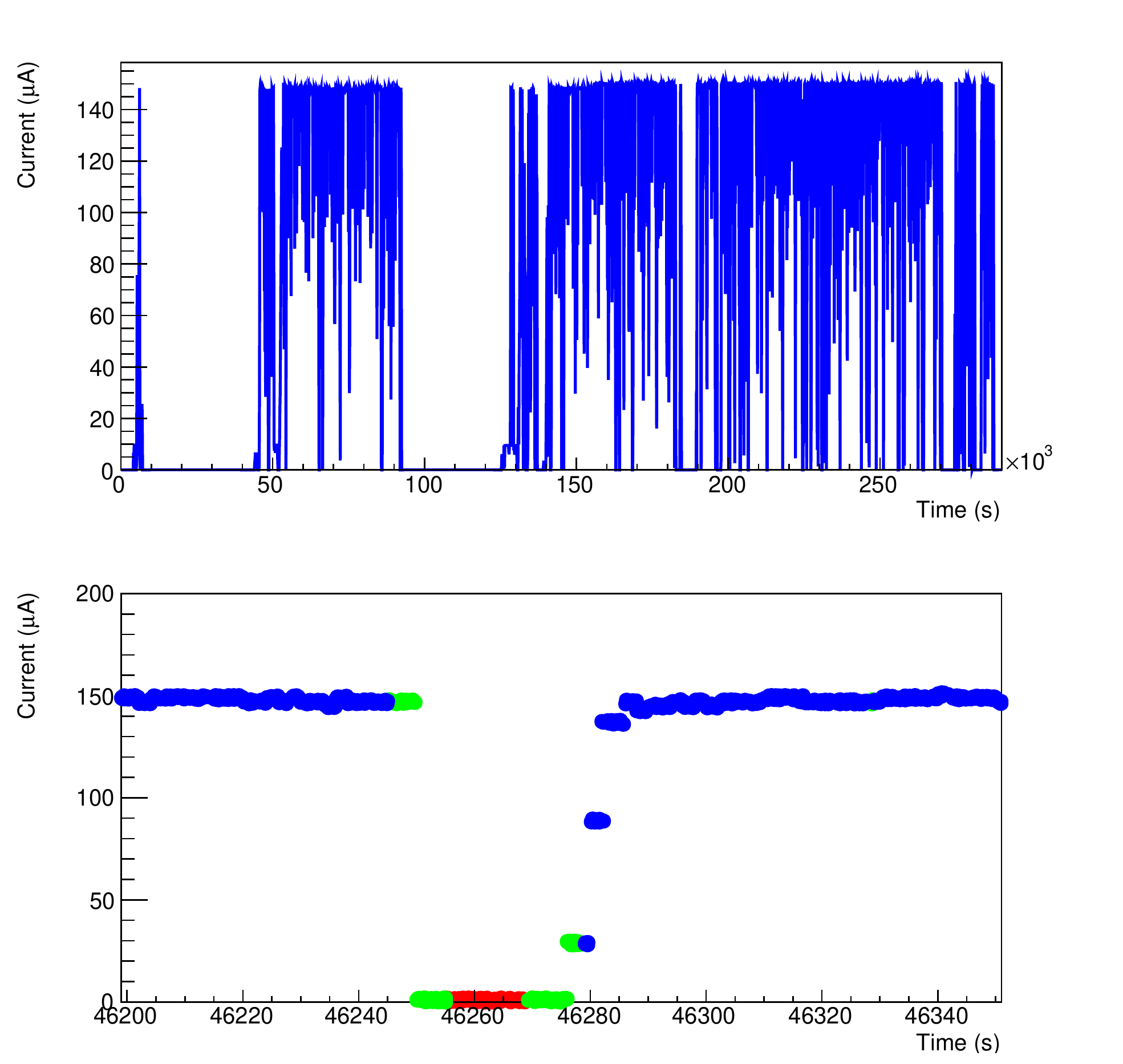}
    \caption{Top panel shows an example of the beam current as a function of time. Bottom panel shows a zoom of a short period: the blue points represent the beam on periods, the red ones are the beam-off periods and the green ones are the periods excluded from the data analysis - see text for more details. (Color online). }
    \label{fig:current}
\end{figure}

Cosmic-ray data were collected before and after the accelerator operations. In addition, during beam operations the beam current was used to identify beam-on and beam-off periods. Events collected 5\,s before and after the beam-trip-time intervals were excluded from the data analysis.
As an example, the beam current as a function of time is shown in Fig.\,\ref{fig:current}.

\subsection{Event reconstruction}

The data reconstruction software was developed using the ``JLab Data Analysis Framework'' (JANA)~\cite{jana}. For each event, the recorded waveforms were processed in order to extract the corresponding charge, amplitude and time. This information was converted into physical units by applying proper calibration constants, and finally written to ROOT-based Data Summary Tapes (DSTs), which were then subsequently used for our data analysis. 
For events featuring a large energy deposition in the crystals with no activity in the surrounding veto systems, the raw waveforms were also saved in the DST, to allow for an event-by-event scrutiny.

For the PbWO$_4$ crystals, each waveform was first corrected, event by event, subtracting the pedestal, which was evaluated as the average value of the first 30 measured samples. The corrected waveform was then numerically integrated within a 160\,ns time window, starting 20\,ns before the maximum. Finally, the integrated charge was converted to MeV units by using calibration constants obtained with beam-produced muons.

\begin{figure}[tpb]
    \centering
    \includegraphics[width=0.45\textwidth]{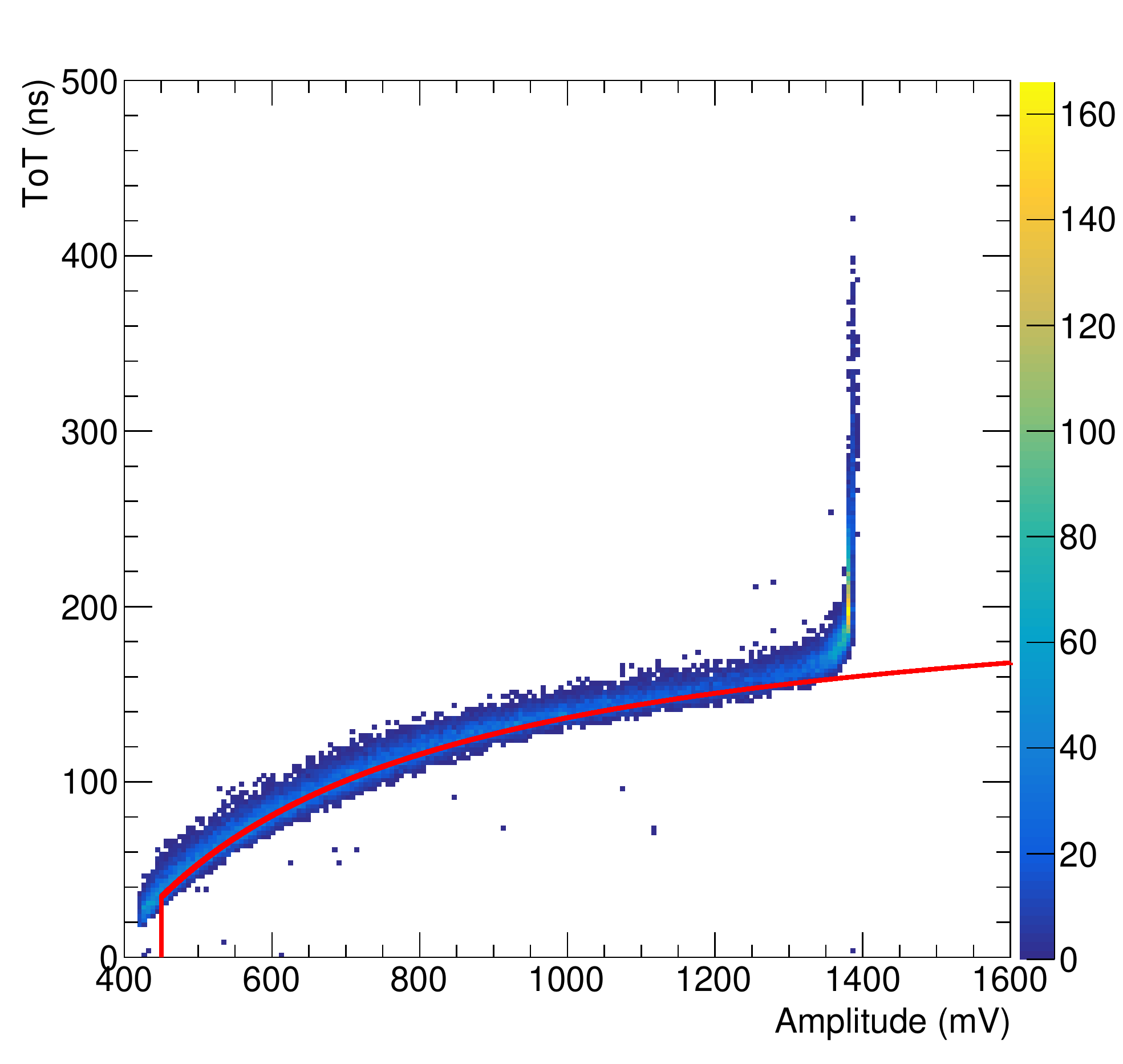}
    \caption{Correlation between the pulse amplitude and Time-over-Threshold value for a single SiPM. The red curve corresponds to the function described in the text. (Color online).}
    \label{fig:ToT}
\end{figure}

For plastic scintillator detectors, we used the amplitude of each signal with respect to the pedestal and its corresponding time. The amplitude was then normalized to the single photoelectron value, as determined during dedicated, random-trigger, runs. Since the main goal of the veto detectors was to unambiguously detect all signals inside the active volume, the gain of the SiPM amplifiers was set to be sensitive to individual photoelectron signals, at the price of a smaller dynamic range that was limited by the amplifier saturation by high-energy signals.
During the calibration phase, however, it was found that, for a better comparison between data and MC, a proper measurement of high-energy events was required. This was obtained by means of the Time-over-Threshold (ToT) technique, applied to the digitized waveforms.
The ToT technique measures the time interval between the leading edge and the trailing edge of a pulse, both determined at the same amplitude threshold. As long as the dynamics of the amplification chain remains linear, the ToT can be used to calculate the corresponding pulse amplitude, even for very large signals. Figure~\ref{fig:ToT} shows, for a single sensor, the correlation between the $ToT$ value (in ns, $y$-axis) vs the pulse amplitude $A$ (in mV, $x$-axis) for a ToT threshold $V^{TOT}_{thr}$ of 400 mV, corresponding approximately to 40\,phe. High-energy events resulting in amplifier saturation are clearly visible. The red curve corresponds to the function\footnote{This function corresponds to the exact $ToT$ vs $A$ relationship for an ideal triangular signal, with $T_0$ being the total signal width.}
\begin{equation}
    A = V^{TOT}_{thr} \cdot \frac{T_0}{T_0 - ToT},
\end{equation}
where the parameters $T_0$ and $V^{TOT}_{thr}$ were determined by a fit to the profile histogram of data in Fig.\,\ref{fig:ToT}, excluding saturation.
The ToT technique was implemented in the reconstruction framework for all sensors. The function parameters were determined separately for each channel.

\subsection{Energy calibration}\label{EnergyCalib}

The calibration was determined with muons produced by an electron beam with an energy of 10.38\,GeV by comparing their measured and simulated energy spectra.
Specific muon trajectories were used as described in Section 4.2.1. The calibration procedure is detailed in Section 4.2.2, and studies of the calibration stability are reported in Section 4.2.3.

\subsubsection{Muon trajectories and detector alignment}
\begin{figure}[t]
    \centering
    \includegraphics[width=.45\textwidth]{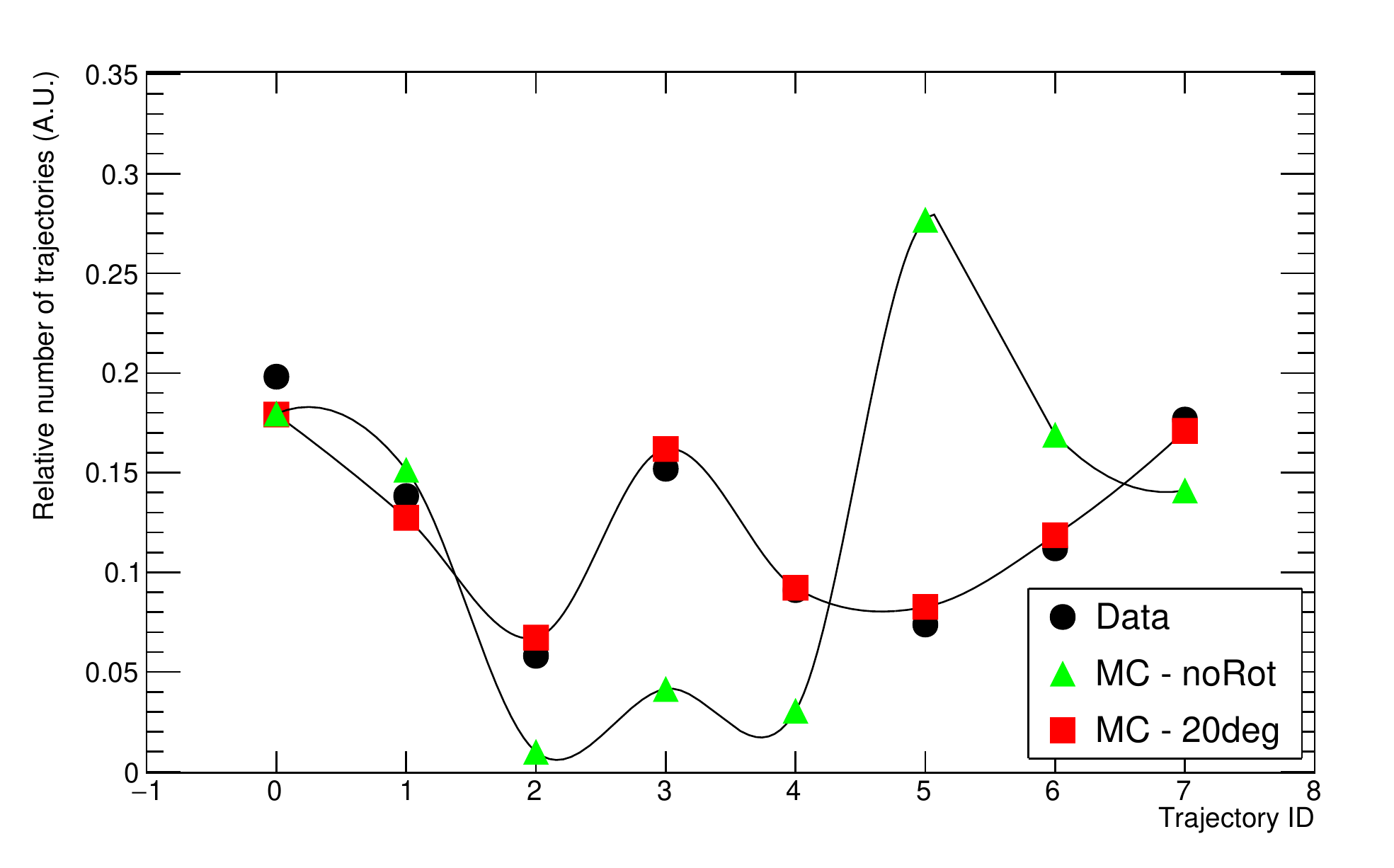}
    
     \includegraphics[width=.45\textwidth]{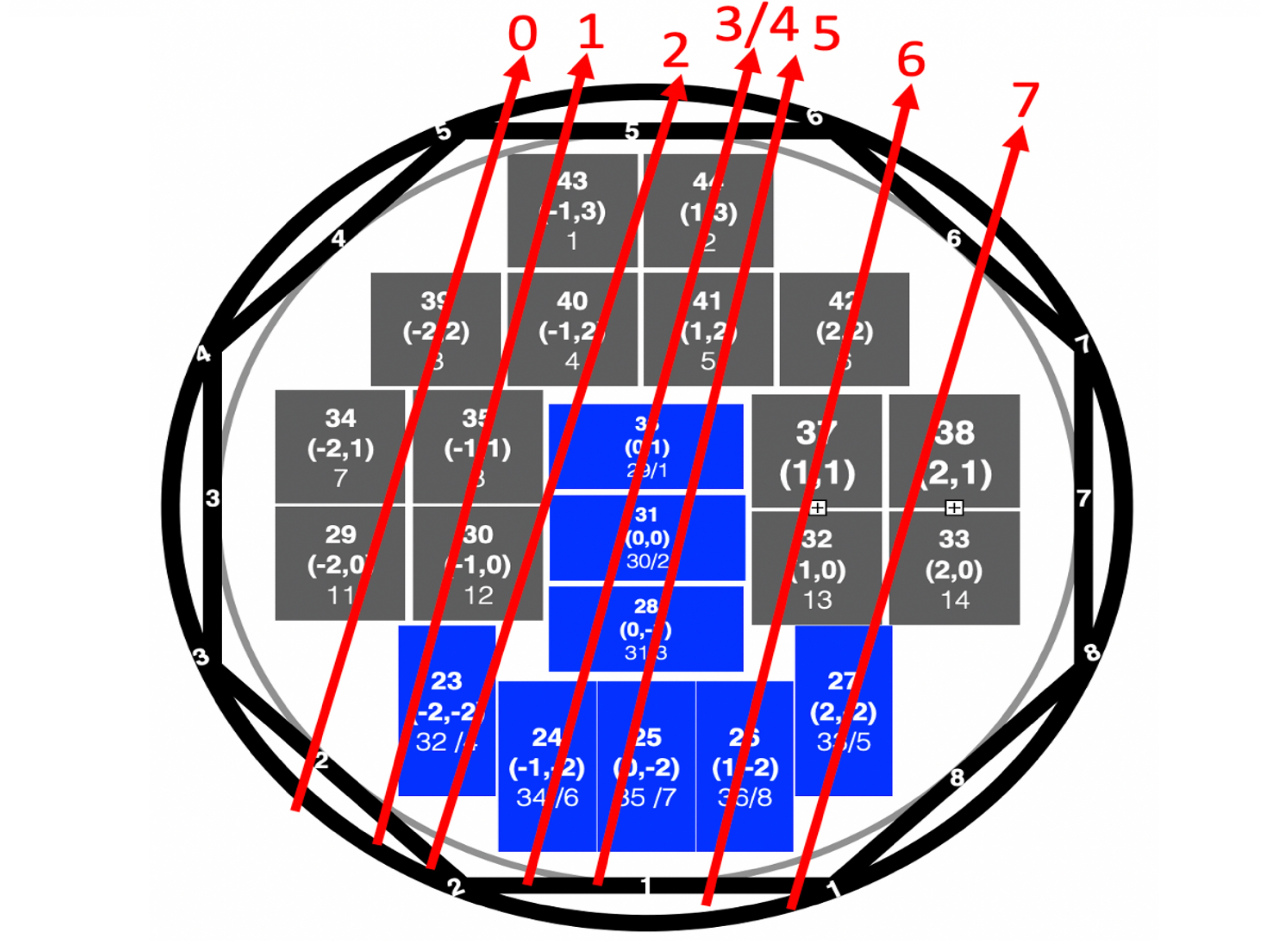}
    \caption{
    Top: relative number of muon trajectories as a function of their ID number for the Bottom half of the ECal. Data are compared with MC predictions for the detector in the nominal position (noRot) and rotated clockwise by 20 degrees. Similar results have been found for the Top half. Bottom: arrows schematically indicate the set of muon trajectories (numbered from 0 to 7) used to calibrate the crystals. Crystal numbering refers to the Bottom half of the ECal. (Color online) }
    \label{fig:Trj} 
\end{figure}

Sixteen sets of muon trajectories were used for energy calibrations: eight for the top half of the ECal and eight for the bottom. They are schematically shown in Fig.~\ref{fig:Trj}. Each trajectory set is defined by requiring the presence of a signal with $\geq$10\,MeV in the relevant crystals (e.g. crystals numbers 29, 34, 39 for the trajectory number 0 of the bottom half). By comparing the rate of events measured for each set of trajectories with that predicted by MC, we discovered that the detector was rotated with respect to beam direction by 20 degrees, as shown in Fig.~\ref{fig:Trj}. The rates of the rotated detector match the MC expectation.

\subsubsection{Calibration procedure}
\begin{figure}[t]
    \centering
    \includegraphics[width=.45\textwidth]{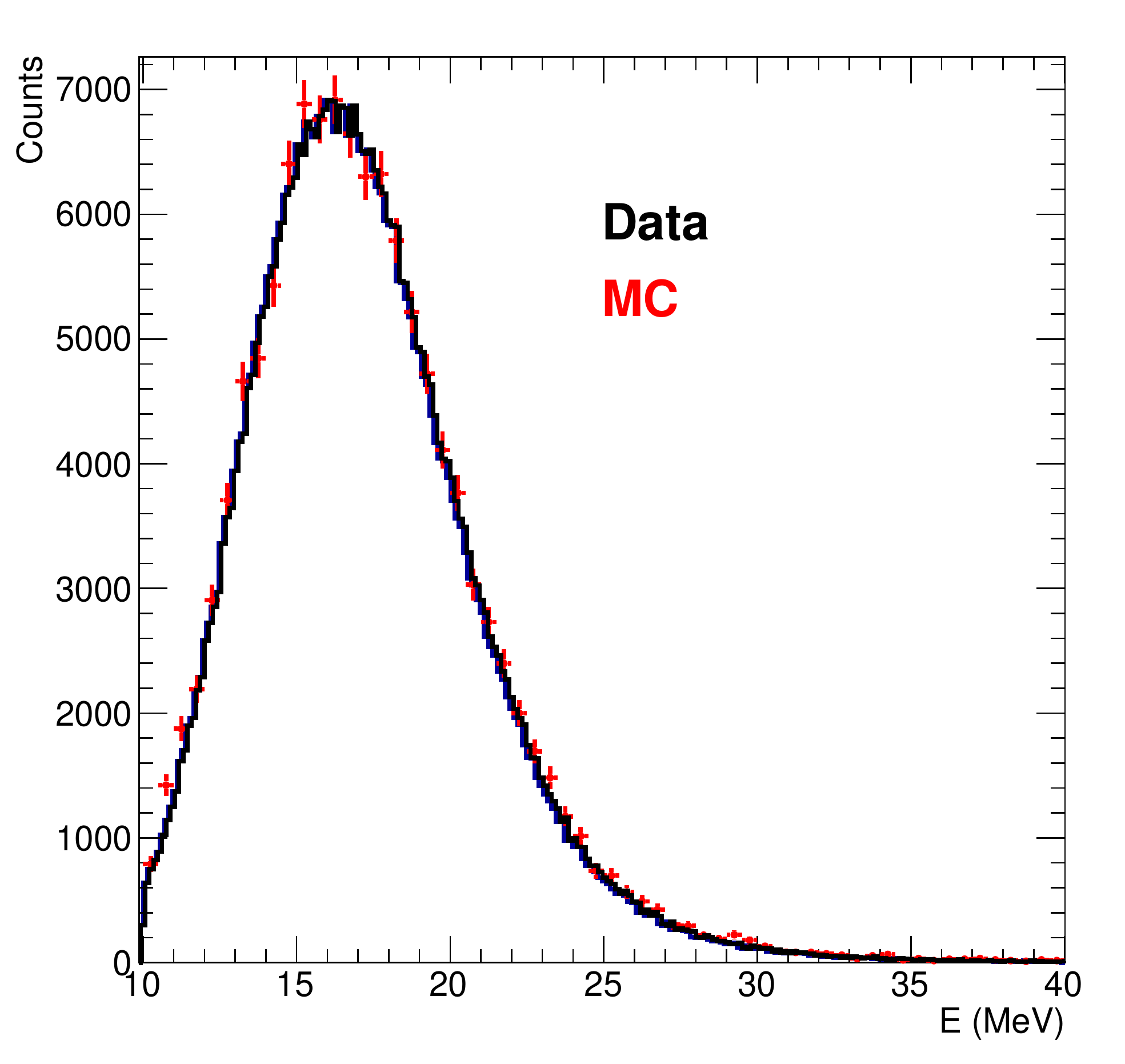}
    \caption{A calibrated energy spectrum of a crystal of the ECal for data at 10.381\,GeV. Red points are for beam-related muons extracted from MC simulations. (Color online).}
    \label{fig:CalibCrys9}
\end{figure}

For each crystal we extracted the energy spectrum of muons crossing the crystal and belonging to the corresponding trajectory set (e.g. TrjID 0 for crystal number 34). 
The same energy spectra have been constructed using MC simulations with BDX-MINI rotated by 20 degrees. 
From each simulated histogram a probability density function (pdf) was extracted. The pdf of each detector was then used to fit the experimental spectrum and extract the calibration coefficient. Fig.~\ref{fig:CalibCrys9} shows an example of the MC simulation superimposed on the calibrated energy spectrum.

\subsubsection{Calibration stability}
\begin{figure}
    \centering
    \includegraphics[width=.45\textwidth]{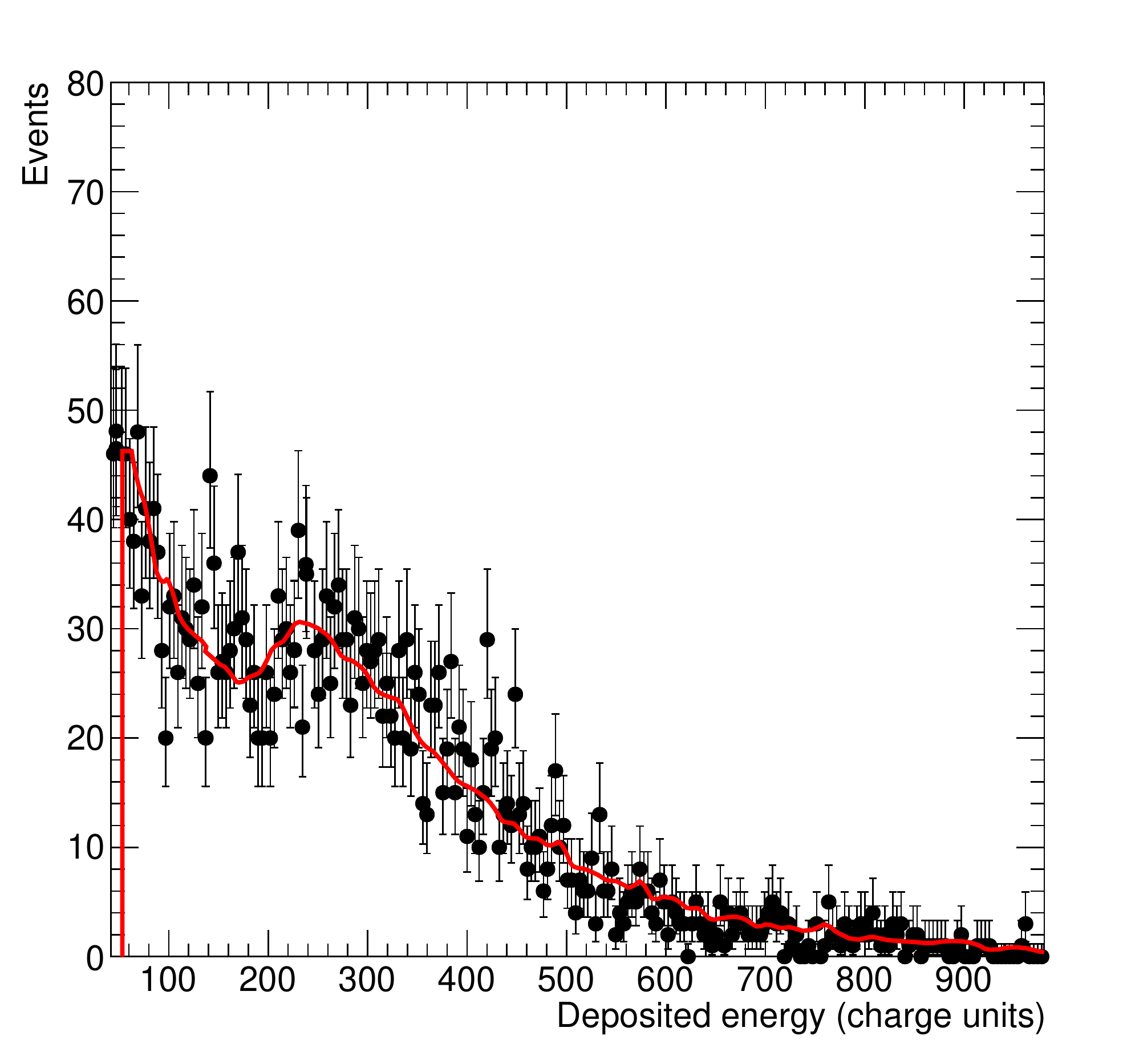}
    \caption{Measured energy distribution for a single ECal crystal, in charge units, for cosmic-rays events selected according to the algorithm described in the text. The red curve represents the result of the fit performed using the template shape obtained from Monte Carlo simulations. (Color online).}
    \label{fig:cosmicsEdep}
\end{figure}

The high rate of muons at a beam energy of about 10\,GeV allowed us to collect enough statistics for our purposes in few hours. Due to the scheduling of the CEBAF accelerator, no other 10\,GeV runs were taken during the rest of the measurement. Therefore, in order to check the stability of the PbWO$_4$ crystal response for the whole BDX-MINI run period, which lasted for several months, we developed an ad-hoc procedure based on cosmic rays.
We defined for each crystal an event selection algorithm resulting in a deposited energy distribution, in charge units, with a clear minimum-ionizing Landau peak. Events were selected according to a combination of the OV and IV SiPMs with the largest signal amplitude. For each crystal, depending on its position in the detector assembly, a different combination was used, based on the photo-detectors closer to it. The additional requirement of a dual coincidence between the IV and OV upper (lower) caps for crystals in the upper (lower) layer selected events corresponding to the passage of cosmic muons with well identified trajectories. A typical example of the energy deposition distribution obtained by this procedure for a ECal crystal is shown in Fig.~\ref{fig:cosmicsEdep}. The same selection procedure was applied to Monte Carlo events, generated according to the procedure described in Sec.~\ref{sec:MC_cosmic_rays}. The resulting MC energy spectrum was then used to develop a template for fits to the measured charge distribution, where the calibration constant (charge-to-energy) was the single fit parameter. To improve the convergence of the fit, only the energy range close to the Landau peak was considered.

\begin{figure}
    \centering
    \includegraphics[width=.45\textwidth]{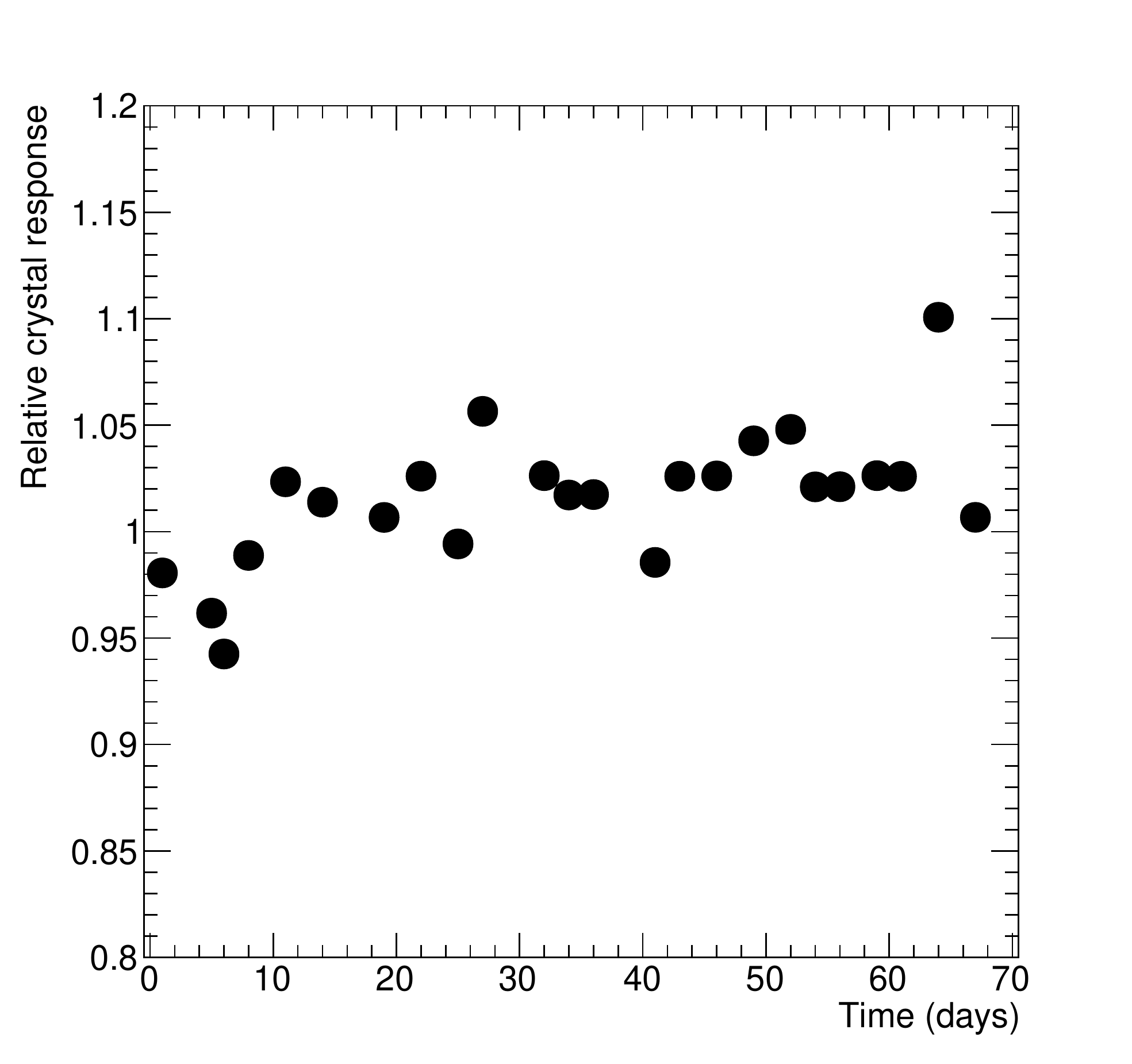}
    \caption{Long term stability of a single crystal over a period of 2 months.}
    \label{fig:stability}
\end{figure}

In order to pinpoint any variations over time of the calibration constants, this procedure was repeated independently for all crystals for each data-taking run. An example of the stability of calibration constant as a function of time (number of data taking days)  for one crystal is shown in Fig.\,\ref{fig:stability}. The points represent the ratio of the calibration coefficients using cosmics during the run and the value obtained with beam-muons prior to the run period. The result indicates that the system was stable to within 10$\%$ over the data-taking period. It is worth noting that this variation also includes the uncertainty in the cosmic calibration procedure challenged by broad MIP peaks and limited statistics. 

\subsection{Veto efficiency}
The overall capability of the two veto systems to reject cosmic rays for our dedicated search for LDM is discussed below. Specific studies on the detection efficiency of the IV Octagon (IV-O) and OV Cylinder (OV-C) were performed using the 10.38\,GeV data. For central muon trajectories, i.e. TrjID\,=\,4 in Fig.\,\ref{fig:Trj}, the measured detection efficiency is compatible with 100\% within the statistical errors ($<0.1\%$). This result is obtained by requiring that at least one SiPM of the IV-O (OV-C) detects a signal that is $\geq$4\,p.e. The use of higher thresholds does not affect the efficiency until they exceed 10\,p.e. A 100\% detection efficiency is also observed by selecting the other trajectories, from TrID=1 to TrjID=7, both in the upper and lower half of the ECal. 

We used selected cosmic ray trajectories to study the detection efficiency of the IV and OV Caps since beam muons had parallel trajectories to them. Central and nearly-perpen\-dicular cosmic-ray tracks were selected by requiring a significant release of energy ($>$10\,MeV) in at least one of the central crystals of the ECal Top half (CryID=6, 8, 9, 10, 13, 14, 15), in the corresponding crystals of the bottom half (CryID=28, 30, 31, 32, 35, 36, 37), and in the Caps that are not under study  (e.g. ECal crystals in coincidence with the OV Caps, OVThr$>$5p.e., when the efficiency of the IV Caps is measured, and viceversa). No activity in all the other crystals and Veto channels is also requested. As for the IV-O and the OV-C discussed above, the measured efficiency of the Caps is 100\% within the statistical errors. 
\begin{figure}
    \centering
    \includegraphics[width=.45\textwidth]{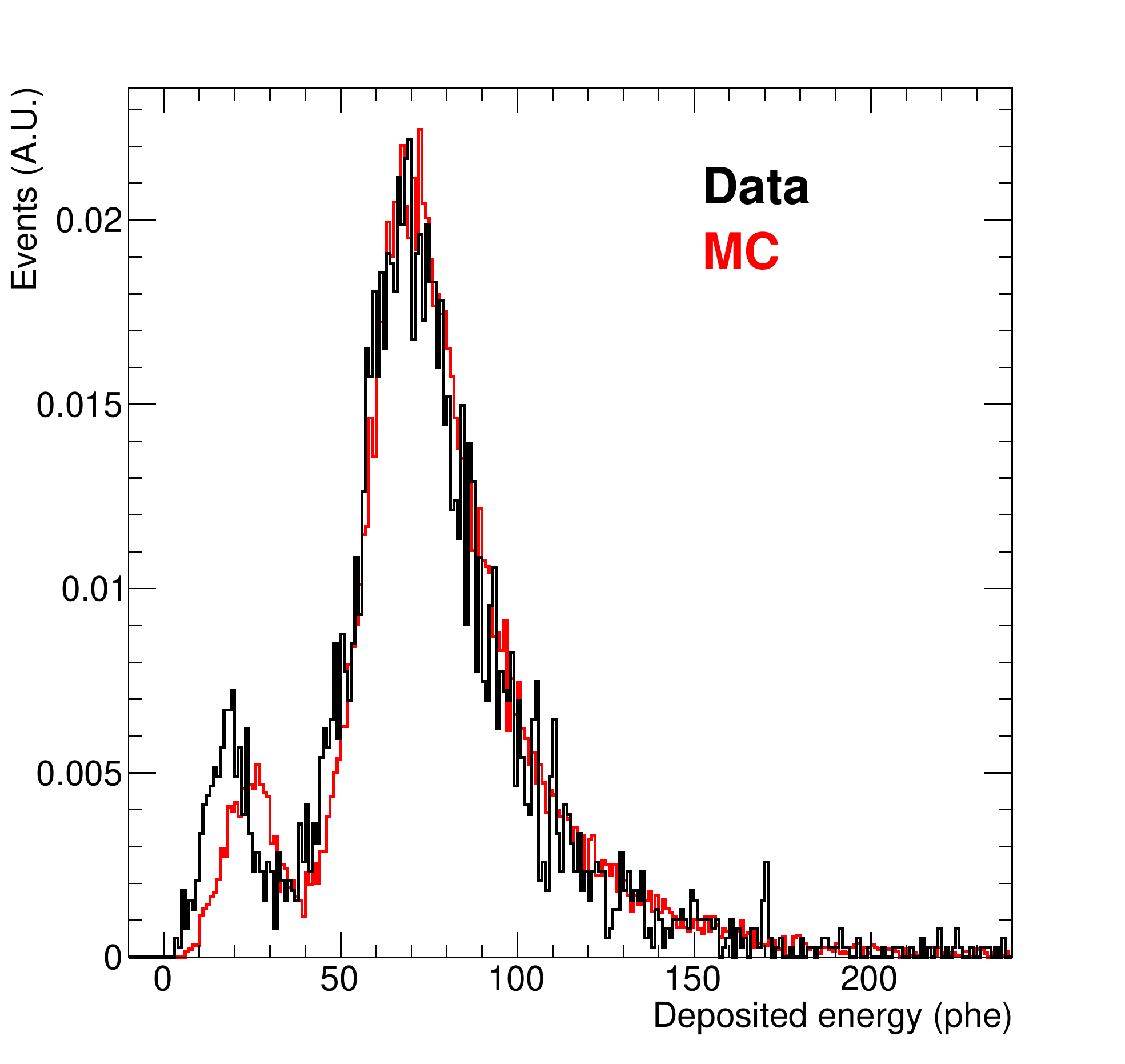}
    \caption{Comparison between data and MC simulation for the OV. The distributions show the response of SiPM\#4 to muons traversing the detector with the trajectory number 0. The two peaks correspond to different muon topologies. (Color online).\label{fig:OVch4tr0}}
\end{figure}

\subsection{Validation of MC description of IV and OV}

The data collected with beam muons were used to check the veto response in the simulation as a function of the position of the hit. As described in Sec. 3.1, the  light transmission in the IV and OV lateral surfaces results in a nontrivial dependence of the signal measured by the SiPMs on the muon hit position on the vetoes. To include this behavior in the MC framework, an effective formula describing light transmission has been obtained from the study of cosmic muons. The muon flux generated by the 10.38 GeV electron beam  provides an independent check of this modeling. A full simulation of the secondary muon beam was performed, and the signal spectra of the different SiPMs of the IV and OV lateral surfaces was compared to the measured distributions, for the different muon trajectories selected with the procedure described in Sec. 4.23. As an example, Fig.~\ref{fig:OVch4tr0} shows the data-MC comparison for the signal amplitude distribution of the SiPM 4 of the OV, for muons crossing the detector with the trajectory 0. The spectrum features a rich structure, with two peaks corresponding to different muon topologies. The MC spectrum reproduces this structure with a reasonable agreement: the main peak is correctly reproduced, while the simulation slightly overestimates the position of the lower one. This level of agreement is typically obtained for all other SiPMs and trajectories of the IV and OV. Given that the vetoes were used only to reject cosmics and not to measure the energy released in the hit, we consider this result quite satisfactory.  

\subsection{Cosmic rejection studies}

\begin{figure}
    \centering
    \includegraphics[width=.45\textwidth]{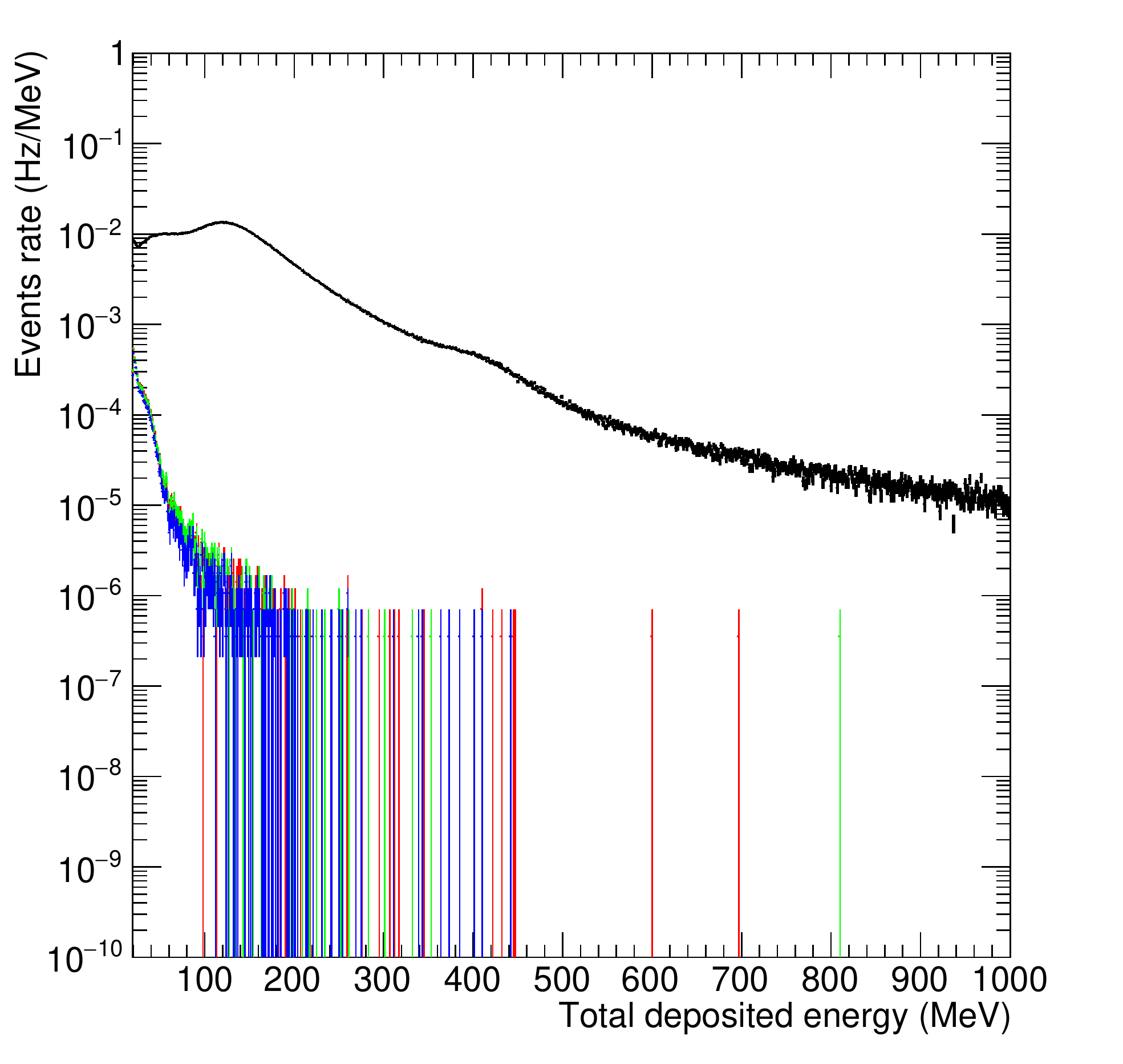}
    \caption{Distribution of the total energy released in the ECal. Different colors refer to the different anti-coincidence selections: black - all events, red OV-anti-coincidence, green IV-anti-coincidence, blue  both veto systems. (Color online).
    \label{fig:spectra_Etot}}
\end{figure}

In order to estimate the veto system capability to reject cosmic events, the beam-off data were analyzed. The data sample considered consists of about 33 days of data taking.
Fig.\,\ref{fig:spectra_Etot} shows the reconstructed energy distribution of events measured by the BDX-MINI, with no condition on the vetoes (black points) and requiring for the anti-coincidence of the veto system with the IV (red points) and with the OV (blue points). In the data analysis the minimum detectable energy was set to 6\,MeV for each crystal and it is required that at least one crystal exceeds an energy of 20\,MeV. Moreover the following definition was used for the veto anti-coincidence. 
An event is considered vetoed if:
\begin{itemize}
\item At least two SiPM signals of the OV (IV) exceed a 2.5 phe threshold within a time window of 200\,ns centered on the time of OV(IV) SiPM with the highest amplitude. 
\item Otherwise, if the previous condition is not satisfied, at least a single SiPM signal exceeds a higher threshold (5.5 phe).
\end{itemize}
In addition, these conditions  were tested on data collected during a random-trigger run and it resulted in a negligible rate of false positive due to electronic noise.

The results show the high rejection capability of the two veto systems. At the minimum-ionizing peak, centered at $\sim$\,120 MeV, the differential counting rate is suppressed by a factor 3800 when the anti-coincidence is requested. Moreover, no cosmic event with energy higher than 450\,MeV is measured without hits in either veto. It is worth stressing that the two systems, characterized by two different geometries,  show similar cosmic-ray rejection capabilities. Therefore the octagon geometry is preferable over the cylindrical ones, since it is less expensive, simpler to assemble, and eventually easier to repair. 

\section{Summary and conclusions}

BDX-MINI represents the first detector specifically designed for a beam-dump experiment searching for LDM particles with an electron beam. 
This paper describes the BDX-MINI detector and its excellent performance during a long LDM measurement campaign performed at JLab. It is a compact detector composed of a PbWO$_{4}$ electromagnetic calorimeter, surrounded by a layer of tungsten shielding and two hermetic plastic scintillator veto systems.
The system performance was studied using cosmic-ray data collected during beam-off periods and with beam-related muons produced in special runs with an electron beam energy of 10.38\,GeV. For the entire measurement period the detector response was constantly monitored and shown to be  stable.  This is a nontrivial result, considering the long run period that the detector was operating outdoors in a field tent.

The innovative solutions of the detector design were validated.  SiPMs were used as the light sensors for an electromagnetic calorimeter consisting of PbWO$_4$ crystals.  The results demonstrate a light yield of $\sim$\,1\,phe/MeV. Moreover, we found that even connecting the SiPM amplification stage at 8~m from the sensors, the noise was low enough to distinguish the individual photo-electrons signals for all SiPMs.
The use of two veto systems, instead of one, has the clear advantage in increasing the cosmic-ray rejection efficiency. For example, the anti-coincidence of both vetoes allowed us to lower the ECal energy threshold from 700-800\,MeV down to 450\,MeV with no background events observed. The two veto systems have different shapes. However, both of them demonstrated a comparable high background  rejection efficiency, indicating that the octagonal geometry is to be preferred in terms of cost and robustness.

This paper provides a detailed description of the solutions adopted to build a robust Monte Carlo framework. 
The Hall-A beam line, beam dump and passive shielding have been included in the simulations as well as the detector response. Cosmic rays, beam induced muons, and neutrinos have been simulated and propagated up to the detector. The detector responses were used to validate their description in the MC.  The selection of cosmic rays and muon trajectories allowed us to parameterize the light transmission in the veto systems as a function of the hit position on the scintillator, including the effect of light absorption due to the WLS grooves. This Monte Carlo simulation framework will be used in the search for LDM interactions and to determine the exclusion limits obtained from the measured signal rates.

\section*{Acknowledgments}
The authors would like to thank the JLab Directorate and Physics Division, and the Italian Istituto Nazionale di Fisica Nucleare for invaluable support during the entire project.
We would also like to thank the INFN technical staff for the excellent work in constructing the detector, JLab Physics technical staff for careful installation, JLab Facilities for design, survey, and logistical support, JLab Networking/Computing for providing connectivity.
Special thanks to CLAS12 Collaboration and PANDA Collaboration (in particular\\ K.-T.\,Brinkmann and H.-G.\,Zaunick)  for providing the PbWO$_{4}$ crystals, and finally to R. Hatcher for generating the GENIE splines for Tungsten. 
This material is based upon work supported by the U.S. Department of Energy, Office of Science, Office of Nuclear Physics under contract DE-AC05-06OR23177. 

\bibliographystyle{spphys}       

\bibliography{mybibfile}

\end{document}